\documentclass[sn-basic]{sn-jnl}%

\PassOptionsToPackage{unicode}{hyperref}
\PassOptionsToPackage{hyphens}{url}

\usepackage{lmodern}
\usepackage{amssymb,amsmath}
\usepackage{ifxetex,ifluatex}
\ifnum 0\ifxetex 1\fi\ifluatex 1\fi=0 
  \usepackage[T1]{fontenc}
  \usepackage[utf8]{inputenc}
  \usepackage{textcomp} 
\else 
  \usepackage{unicode-math}
  \defaultfontfeatures{Scale=MatchLowercase}
  \defaultfontfeatures[\rmfamily]{Ligatures=TeX,Scale=1}
\fi
\IfFileExists{upquote.sty}{\usepackage{upquote}}{}
\IfFileExists{microtype.sty}{
  \usepackage[]{microtype}
  \UseMicrotypeSet[protrusion]{basicmath} 
}{}
\makeatletter
\@ifundefined{KOMAClassName}{
  \IfFileExists{parskip.sty}{%
    \usepackage{parskip}
  }{
    \setlength{\parindent}{0pt}
    \setlength{\parskip}{6pt plus 2pt minus 1pt}}
}{
  \KOMAoptions{parskip=half}}
\makeatother
\usepackage{xcolor}
\IfFileExists{xurl.sty}{\usepackage{xurl}}{} 
\IfFileExists{bookmark.sty}{\usepackage{bookmark}}{\usepackage{hyperref}}
\hypersetup{
  hidelinks,
  pdfcreator={LaTeX via pandoc}}
\usepackage{longtable,booktabs}
\usepackage{etoolbox}
\makeatletter
\patchcmd\longtable{\par}{\if@noskipsec\mbox{}\fi\par}{}{}
\makeatother
\IfFileExists{footnotehyper.sty}{\usepackage{footnotehyper}}{\usepackage{footnote}}
\makesavenoteenv{longtable}
\usepackage{wrapfig}
\usepackage{rotating}
\usepackage{epstopdf}
\usepackage{graphicx}
\usepackage{pdflscape}
\usepackage{url}

\title[Faint Meteor Detector Using Matched Filtering]{Development of a Very Faint Meteor Detection System based on an EMCCD Sensor and Matched Filter Processing}

\author[1]{P. Gural}\email{pgural@gmail.com}
\author[2,3]{T. Mills}
\author[2,3]{M. Mazur}
\author[2,3]{P. Brown}
\affil[1]{Gural Software and Analysis LLC, Lovettsville, VA 20180 USA}%
\affil[2]{Department of Physics and Astronomy, University of Western Ontario, London, ON, N6A 3K7 Canada}
\affil[3]{Institute for Earth and Space Exploration (IESX), The University of Western Ontario, London, Ontario N6A 3K7, Canada}
\jyear{\year}

\usepackage{aas_macros}
\usepackage{hyperref} 
\hypersetup{colorlinks,citecolor=blue,linkcolor=blue,urlcolor=blue}

\begin{document}


\abstract{The mass ranges of meteors, imaged by electro-optical (EO) cameras and backscatter radar receivers, for the most part do not overlap. Typical EO systems detect meteoroid masses down to 10$^{-5}$ kg or roughly magnitude +2 meteors when using moderate field of view optics, un-intensified optical components, and meteor entry velocities around 45 km/sec. This is near the high end of the mass range of typical meteor radar observations. Having the same mass meteor measured by different sensor wavelength bands would be a benefit in terms of calibrating mass estimations for both EO and radar. To that end, the University of Western Ontario (UWO) has acquired and deployed a very low light imaging system based on an electron-multiplying CCD camera technology. This embeds a very low noise, per pixel intensifier chip in a cooled camera setup with various options for frame rate, region of interest and binning. The EO system of optics and sensor was optimally configured to collect 32 frames per second in a square field of view 14.7 degrees on a side, achieving a single-frame stellar limiting magnitude of m$_G$ = +10.5. The system typically observes meteors of +6.5. Given this hardware configuration, we successfully met the challenges associated with the development of robust image processing algorithms, resulting in a new end-to-end processing pipeline now in operation since 2017. A key development in this pipeline has been the first true application of matched filter processing to process the faintest meteors possible in the EMCCD system while also yielding high quality automated metric measurements of meteor focal plane positions. With pairs of EMCCD systems deployed at two sites, triangulation and high accuracy orbits are one of the many products being generated by this system. These measurements will be coupled to observations from the Canadian Meteor Orbit Radar (CMOR) used for meteor plasma characterization and the Canadian Automated Meteor Observatory (CAMO) high resolution mirror tracking system.} 


\maketitle

\section{INTRODUCTION }
\label{sec:intro}

Most of the annual mass of extraterrestrial material accreting onto the Earth is from meteoroids in the mass range of 10$^{-8}$ kg \citep{Love1993a} and larger. However, as mass averaged meteoroid impact speeds are in the 15-20 km/s range \citep{Moorhead2017}, meteoroids near the mass influx peak produce very faint meteors, typically near +12 visual magnitude. Hence to begin to probe the meteoroid range where most mass accumulates to Earth, it is necessary to push detection limits down to ever fainter meteor magnitudes. 

While radar systems can probe this mass range \citep{Baggaley2002}, radar suffers from several biases (such as initial radius and low ionization probability at slow speeds) which makes inferences about the source population from radar difficult. Optical detection of very faint meteors is technologically challenging. The short duration of meteors and their high angular velocity leads to rapid motion and trailing losses across an image plane. As it is desirable to collect as many photons as possible, comparatively large optical apertures have previously been used for this purpose \citep{Cook1980, Pawlowski2001, Kaiser2005, Iye2007, Ohsawa2018, Ohsawa2020}.

However, these large apertures tend to restrict the field of view, even for very fast optics, making trailing losses even more significant. The net result is that even with ultra-sensitive, low noise modern detectors and large telescopes, optical meteor studies below +8 magnitude are sparse. Moreover, by the very nature of the large telescopes used in past studies, almost all measurements have been single station. No multi-site survey of faint meteors has been performed, which would allow heights, velocities, and orbits with a sufficient population sampling, so as to be compared with ongoing video network surveys performing collections and analysis on larger mass meteoroids. 

To address the need for a multi-station survey of the faint meteor population, we have deployed highly sensitive single photon cameras consisting of an Electron Multiplied CCD (EMCCD) coupled together with automated meteor detection software operating near the noise limit of the sensors. The goals of this survey are fourfold:
\begin{enumerate}
    \item To coincidentally detect the optical counterpart to underdense specular echoes recorded by the Canadian Meteor Orbit Radar (CMOR), thus enabling better calibration of the radar mass scale.
    \item To perform a long-term survey of the faint meteor population covering all speeds from 12.3 – 71.9 km/s with corresponding limiting mass in the 10$^{-4}$ – 10$^{-8}$ kg range (optical magnitude range +0 to +8) with emphasis on population physical and orbital characteristics. This will allow comparison to dynamical models of the meteoroid population. A particular goal for the survey is to quantify the origin and flux of the low velocity (<20 km/s) meteoroid population.
    \item To measure the flux and mass distribution of both meteor showers and the sporadic complex at small enough masses to compare with measurements from other techniques (e.g. radar).
    \item Achieve goals 1 through 3 by obtaining precise and automated measurements of each detected event on a per frame basis that rely on both new innovations and well-established astrometric and photometric techniques
\end{enumerate}

As the first goal requires the detection of a meteor’s optical light curve below the underdense radar limit of 10$^{-7}$ kg or approximately +6 in visual magnitude  \citep{WerykBrown2012}, the instruments must detect a significant fraction of all meteors with peak magnitudes at or below the +6 limit. In particular, as the Canadian Meteor Orbit Radar (CMOR) has an absolute detection limit near meteor magnitude +8, it would be desirable that the system detect light curves down to this limiting absolute magnitude. If achievable, some or all of the specular meteors detected by CMOR will be visible to the optical system. This goal of working in the mass range associated between +5 to +8 meteors is the primary driver of our system’s minimum sensitivity requirement. 

As previous multi-site optical surveys have been magnitude limited to +3 to +4 \citep[e.g.][]{Jenniskens2011a, Koten2019} collecting significant numbers of meteors down to a limiting peak absolute magnitude of +6.5, would provide roughly an order of magnitude deeper sensitivity in mass than prior surveys based on the power law distribution of meteor counts with mass. Note that we do not require our survey to be complete to this magnitude, (our survey is complete to about +5), but to collect sufficient meteor numbers to +6.5 such that corrections for detection efficiency and geometry can still produce statistically meaningful results.

In addition to the above experiment driven requirements, for both practical reasons and based on prior experience in automated meteor detection, the system development specifications began with a desire to detect and obtain metric data down to SNR = 2 dB with no more than a 3:1 false-to-true ratio (SNR is defined as 10 log$_{10}$ ( peak pixel signal / standard deviation of the noise background ). A secondary requirement was that the processing architecture of hardware and software be designed such that one night of data could be completely processed before the next night’s data collection, to avoid race conditions and possible data loss. 

This paper focuses mainly on the algorithmic development details and the associated software performance with regard to meeting the survey specification sensitivity.  It starts with a brief description of the hardware acquired and site deployment to provide context with respect to the software implementation chosen. Given the sensor’s properties and operational parameters, a characterization of meteors as realized on the EMCCD focal plane is also detailed. This is followed by a description of the core processing algorithm that utilizes matched filtering for three distinct phases: meteor detection, false alarm mitigation, and measurement refinement. A step-by-step walk-through of the entire processing pipeline is described, followed by performance evaluations on both simulated and real imagery. This includes a first look at directly applying a full-blown matched filter as a front-end detector and how that compares in performance to the more run time efficient, cued matched filter approach as implemented in the operational pipeline. 

A more complete hardware summary and debiasing study of the early data will be provided in another paper (Mazur in preparation, 2022) along with a preliminary study of meteor properties from this system \citep{Mills2021}.

\section{HARDWARE DESIGN AND DEPLOYMENT}
For our faint meteor survey, the set of primary requirements drives one towards an intensified sensor with fast imaging lens and somewhat narrow field of view, where the latter involves a trade-off of limiting magnitude, trailing losses, imaging aberrations, and sky coverage. For the sensor we chose a relatively new low-light level camera system developed by NuVu Cameras referred to as an electron-multiplying CCD (EMCCD). This had been developed explicitly for extremely low-light video imaging situations requiring a very low background noise level \citep{Daigle2009}. The Nuvu EMCCD is a Peltier cooled system that operates in our setup at -60 degrees C, provides a thermal noise level below 0.01 e/second/pixel, and readout noise comparable to this level of dark current. 

The particular camera used for our survey is the Hnu1024HS  whose Teledyne e2v CCD201-20 sensor has a 13$\mu$ pitch focal plane comprised of 1K x 1K pixels with 16-bit dynamic range. Full-frames can be read out at 16.7 fps but we operate with 2x2 binning and a frame rate of 32 fps, thus effectively having more measurement points along the meteor track and reduced trailing losses. 

For the survey we have deployed four of these cameras, two at The University of Western Ontario’s Elginfield field station and two at the Canadian Meteor Orbit Radar site near Tavistock, Ontario. The installation of the EMCCD instrument suite was able to take advantage of the infrastructure already in place for several other meteor collection systems resident in the CAMO sheds at these two sites. In particular, these roll-off roof stations are entirely automated, simplifying daily EMCCD operations and permitting remote access to processed results. 
The four cameras are configured as two pointing pairs as seen in the photo of Figure 1. 

\begin{figure*}
\begin{center}
\includegraphics[width=\linewidth]{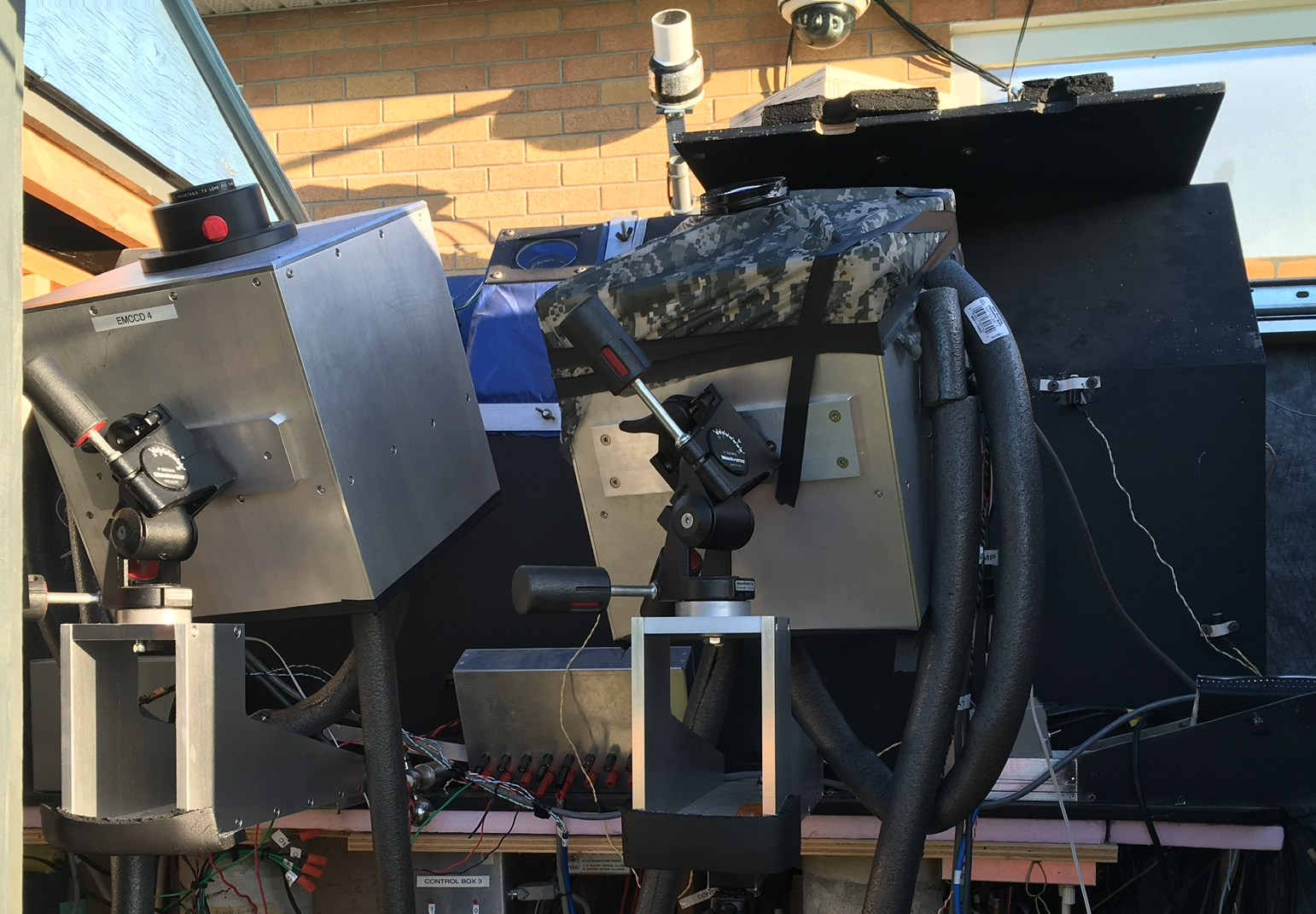}
\caption{ EMCCD system at the Elginfield CAMO station. EMCCDs are the proximal box-like units.}
 \label{emccd}
\end{center}
\end{figure*}

One camera of each dual-site pair is oriented at a higher altitude angle and thus optimized to capture meteors ablating above 100 km, while the second pair pointing at a lower altitude has been optimized for a more restricted height range of 70 to 120 km. The trade-off is that it while the higher pointing camera pair possesses slightly higher sensitivity than the low altitude pair due to the higher camera's closer range to the meteors, the lower pair detects more meteors due to its significantly larger collection area intersecting the height range where meteors ablate. The cameras are pointed in a northerly direction overlapping other CAMO instrumentation with a slight offset from due north to allow for stellar field drift across all the pixels. This pointing attitude permits updates of the flat field using a stare mode collection. Operations are limited to the darker periods of lunar phase and elevation when the ground illuminance is below 0.005 lux (see page 493 of \citet{Seidelmann1992}).

\begin{figure*}
\begin{center}
\includegraphics[width=\linewidth]{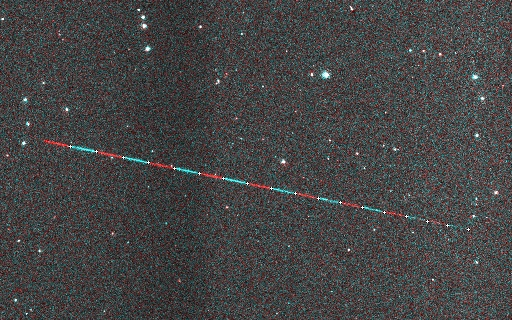}
\caption{Single stacked frame image from the NuVu EMCCD operated in 2x2 binning at 32 fps. Meteor shown as a temporal maximum image stack of alternating frames mapped to red and cyan, with leading edge pick points highlighted as white crosses.}
 \label{stack}
\end{center}
\end{figure*}

For most of the duration of this survey, which began in 2017, all cameras use a 50mm f/1.2 Nikkor lens, with a resultant square field of view (FOV) of 14.7 degrees, giving an angular resolution of 1.72 arcminutes per pixel. The camera’s current configuration settings use an electron multiplying gain of 200, which was found to be optimal for meteor detection but with the drawback that pixel saturation occurs near 38,000 digital units (DU). Thus in operational mode the systems essentially are using only 15 bits of dynamic range out of the available 16 bits. Figure \ref{stack} shows a multi-frame image stack with a meteor from one of the cameras. The low-light performance of a single frame when operated with the 31.25 msec exposure time, produces a limiting magnitude for stars of +10.5.

Both the camera control and frame capture ingestion are handled by a Linux based server, which writes and stores raw video frames into a custom “VID” file. Each frame contains a 116-byte long header embedded in the first row of the image, which records various parameter specifications such as image size, bit depth per pixel, quality flags, universal time for the frame, and the sequence number of each image. Each VID file is currently configured to hold 10 minutes of video or 19,200 frames, encompassing 10 Gigabytes of imagery. To avoid frame drops due to the server's heavy collection loads from running multiple EMCCD camera systems, files are processed the following morning for meteor event detection, followed by multi-site triangulation and orbit estimation. 

The resultant meta data is posted automatically to a web accessible server for inspection and analysis. Automated detections, which include both true meteors and false alarms, are extracted and saved in shorter >2 second video files which typically include 1 second of pre-event background-only frames. Each site employs one computer to control a pair of EMCCD cameras as well as manage the circular image buffer's file storage. The disk space available is capable of retaining approximately 60 hours of raw video. As this capacity can span only several days of imagery, these large continuous collection VID files are aged off (deleted) on a first in, first out basis. Candidate detected events are retained and archived indefinitely as snipped out VID files. The storage system uses a RAID array so there has been no data loss resulting from disk errors over the time the systems have been in operation. 

\section{METEOR CHARACTERIZATION – GENERAL CONSIDERATIONS}

For a given sensor system deployed in the field, one must characterize the focal plane response of the expected meteor events to be encountered. Once that is done, the design of an optimal detection solution in software can proceed. This involves characterizing the meteor traces in terms of their potential range of apparent angular velocity and evaluating both the detection limiting magnitude and the thresholds required for screening out other moving object false alarms such as artificial satellites. 

Meteor entry velocities (V$_\infty$) range from 12.3 to 71.9 km/sec, which can be converted to apparent angular velocity using Equation \ref{eq1} for a radiant angular distance “D” and range “q” from the meteor to the observer in km \citep{Gural1999}:

\begin{equation}\label{eq1}
\omega(^\circ/sec) = \frac{180}{\pi} V_{\infty} \frac{sin (D)}{q},
\end{equation}

Four bounding cases span the expected streak lengths of meteors and artificial satellites per frame. In each case listed below, the range “q” is defined by the 45$^{\circ}$ average pointing altitude of the EMCCD cameras with the slowest meteor’s maximum begin height assumed to be 90 km and the fastest meteor’s minimum end height set to 85 km \citep{Ceplecha1998e}. 

While it is possible for meteors to be outside this range of heights, the vast majority of meteors at these speed boundaries for video magnitudes are found within these lower and upper speed bounds as shown in \citet{Jenniskens2016b}. Note that since the slowest meteor can effectively have zero apparent angular velocity for a direct line-of-sight approach angle, the apparent angular speed lower bound will be computed assuming the meteor has at least a 15° radiant distance. This holds for the majority of meteors encountered with our north celestial pointing direction. To characterize a more typical sporadic meteor, in order to highlight how much faster they are relative to our apparent angular speed cutoff, an average radiant distance of 45° from the sporadic sources was used in the computation. 

Using these considerations a representative set of bounding speeds for our cameras include:
\begin{enumerate}
    \item 	$\omega$ = 0.8$^\circ$/sec or 0.9 pixels/frame for a low orbiting satellite (ISS) at 400 km moving 7.7 km/sec
    \item $\omega$= 1.4$^\circ$/sec or 1.6 pixels/frame for a slow meteor at 90 km moving 12.3 km/sec (q=126 km)  
\item $\omega$= 8.0$^\circ$/sec or 8.7 pixels/frame for a typical sporadic at 90 km moving 25 km/sec (q=126 km)  
\item $\omega$= 34.5$^\circ$/sec or 38 pixels/frame for a fast meteor at 85 km moving 71.9 km/sec (q=119 km)  
\end{enumerate}

The slow meteor bound provide a good threshold to cut out any false alarms arising from artificial satellites. Most satellites orbit at a higher altitude than the International Space Station (ISS) and thus will have even slower apparent angular rate than 0.9 pixels/frame. 

From these considerations, for daily operations, a cutoff of 2 pixels/frame was chosen to provide margin when testing against the minimum speed constraint given that the initial velocity estimation has an inherent error bar. This was later validated by examining raw data and ensuring that no artificial satellites were detected.

The fast meteor angular velocity bound impacts the design considerations of the detection algorithm. As discussed later, a clustering algorithm (which we term a blob detector) was chosen over a Hough line detector for best computational speed. However, if the meteor streak lengths are too long, the clustering approach can run into centroiding issues. Thus, a hierarchical pixel aggregation (downscaling) scheme was developed to maintain both the clustering algorithm’s detection performance and retain a high first-pass throughput efficiency of computation.

Another consideration is the angular velocity loss in magnitudes given the angular resolution of each pixel. At our frame rate, no trailing losses occur for motion with apparent angular speeds <0.92$^{\circ}$/sec, corresponding to a visible object dwelling in an EMCCD pixel during the full frame integration time. However, trailing losses happen naturally in meteor detection systems since most meteors have higher apparent angle rates. The effective magnitude loss for the slow meteor bounding speed as defined above is 0.5. However, for a typical sporadic meteor speed of 25 km/sec the loss is 2.3 magnitudes, and for the fastest meteors the loss can approach 4 magnitudes. 

Given the limiting stellar magnitude of +10.5 for the EMCCD cameras, the expected limiting magnitude for typical meteors should be +7.5 to +8.5. In practice, we find that a meteor has to have a peak magnitude above +6.5 to be automatically detected (see later), though the fading portions of the light curve can be measured in some cases to optical magnitudes fainter than +8 before being lost in the background.

While characterizing meteor behavior in our system, we also created a simulation to evaluate both the performance of the detection algorithm and estimate its detection biases. This was done by injecting artificial meteors of fixed brightness into synthetic EMCCD imagery with a flat distribution of focal plane trail lengths. 

\begin{figure*}
\begin{center}
\includegraphics[width=\linewidth]{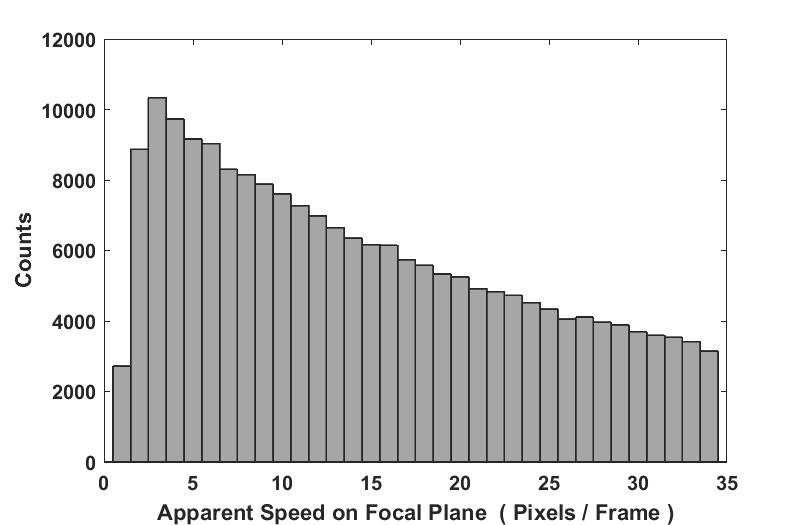}
\caption{Simulation result showing the number of detected meteors from an initial flat velocity distribution (equal number of simulated events in each velocity bin). This assumes a 512 pixel square focal plane, 35 pixels/frame max speed, and 4 frame full meteor segments within the FOV.}
 \label{fig:bias}
\end{center}
\end{figure*}

As shown in Figure \ref{fig:bias}, this basic simulation revealed that a significant selection bias in meteor speeds is present and arises from geometric system constraints, despite a uniform distribution of apparent angular velocities, directions, multi-frame duration, and starting position that were used. The geometric visibility results favored the appearance and detection of preferentially slower meteors with shorter duration. This amounted to almost a threefold increase in the slowest encounter geometries versus the fastest apparent angular velocities across the focal plane (given the EMCCD camera specifications used on this project). There was also a less pronounced ten percent variability favoring angular motions oriented along the diagonals of the rectangular FOV. These effects occur because of the finite size of the camera FOV and a detection requirement to have at least four frames with full streak lengths, reside completely within the FOV. The four frame requirement is needed to ensure the quality of triangulation and orbital parameter estimation when aggregating multiple site measurements together.

Meteors which cross the frame corners and are either entering or exiting the field, or that are faster and thus longer in length per frame have a lower probability of meeting the four-frame requirement unless they are more centrally aligned in the FOV than their shorter/slower counterparts. This bias only addresses the basic geometric selection effects and is by no means a complete picture of the biases present, since there are additional impacts from the mass/magnitude distribution, light curve, detection algorithm biases, and more realistic entry speed distribution and duration of real meteors. But a more in-depth evaluation of that aspect is beyond the scope of this paper and will be addressed in a followup paper (Mazur in preparation, 2022). Suffice it to say that the finite FOV's bias towards slower and thus shorter duration meteors does help to meet one objective of the project favoring the characterizing of the low velocity sporadic population.

\section{MATCHED FILTER DETECTION}

There are many options that can be utilized in the image processing chain for extracting high quality measurement data from multi-frame meteor imagery. The collection and processing steps which can be tuned to improve detection, are broadly categorized as follows:
\begin{itemize}
\item Setting up an optimal collection system configuration (sensor, lens, cadence, pointing, storage)
\item Pre-processing imagery clean-up (darks, flats, hot pixel removal, global background equalization)
\item Stationary background removal (stars, fixed pattern noise) and clutter suppression (read, thermal and Poisson noise)
\item Moving object detection which is tuned towards finding multi-frame propagating meteor streaks
\item Post-detection calculation of meteor positions per frame and their astrometric calibration
\end{itemize}

Details for each of these steps are provided in the next section describing the image processing workflow and key innovations that were developed as part of this study. The current section will focus more broadly on the design trade space for the primary detection algorithm. Note that in Appendix A, a table of alternative processing options for many of the bullet points above is outlined. These had all been under consideration when formulating the initial image processing pipeline.  Both the table in the appendix and the final pipeline work flow are based on years of legacy development work and software implementations that had been previously applied to video meteor detection (for a review see \citet{Koten2019}).

Initially, it was considered that the processing throughput of circa 2017 CPUs and GPUs had reached a point where fully-blind matched filter processing could be applied directly to multi-frame image sequences for the detection of meteors. Since the goal was to detect as faint a meteor as possible, a matched filter was known to be the optimal detector in a Gaussian noise environment \citep{VanTrees2003}. Thus by using a matched filter approach it was hoped to reliably reach down in detection sensitivity to the level of the background noise clutter. 

The terminology fully-blind refers to not knowing a priori, the various motion characterization aspects of

\begin{enumerate}
    \item where in the focal plane the meteor may be located,
    \item at what apparent angular speed and
    \item in which direction the meteor is traveling
\end{enumerate}

The matched filter approach works by hypothesizing all the possible starting positions, realizable speeds, and directions, as it builds and applies numerous multi-frame motion templates on an image sequence. The goal is to find the one motion template with the best spatial-temporal-amplitude alignment with the meteor trail as expressed through the highest maximum likelihood estimate. This involves a template matching operation spanning all meteor frames given the hypothesized motion, which is computationally feasible with present day computers when applied to the somewhat similar problem of multi-frame asteroid motion detection in telescopic imagery \citep{Gural2005, Gural2019}.

However, the number of potential motion hypotheses given the EMCCD pixel scale, a four-frame sliding processing window, the range of apparent angular velocities for meteors, and searching at unity pixel template resolution, is found to require 5000 motion hypotheses applied to each of 512 x 512 pixels requiring 1/4 Tflops/second to operate in real-time. Actual wall-clock timing when using a single-core Ryzen 7 processor, took 21 hours to process 2 seconds of imagery thus requiring 9 months to process a single 10-minute video sequence.

Since the project requirement was to process data at nearly the collection data rate (allowing for one half day latency until the next collection night), using a fully-blind matched filter as a front-end detector was not practical, given the CPU processors available to the project during development. It should be noted that high end GPUs are now operating at the Tflop level and a first pass matched filter detector could be revisited as an option in the future. In tests it was found that a fully matched-filter based detection approach had three times higher detection rate than our algorithm (discussed later in the performance section where such an approach was applied and tested on a limited data set).

Given the computational load requirements and limitations, a hybrid algorithmic approach was examined leveraging an early automated meteor detection application called MeteorScan \citep{Molau2005}. The core concept in the initial design of MeteorScan was to use a very fast front-end detection process with potentially high false alarm rate, but good faint meteor detection sensitivity. A streak detector was used based on a multi-frame Hough transform with a low enough threshold to effectively perform detection near the imagery noise floor. Potential detection candidates would then be re-evaluated with a more robust matched filter approximation algorithm (simplified for computational speed),  removing false alarms and refining positional estimates. 

In this hybrid detection approach, the initial real-time front-end detection process substantially reduces the set of motion hypothesis templates to be searched. Thus the potential detection candidates would only occasionally feed the more computationally costly matched filter, making run times much shorter. A hybrid design philosophy that was ultimately chosen for this project's EMCCD processing pipeline.

\begin{figure*}
\begin{center}
\includegraphics[width=\linewidth]{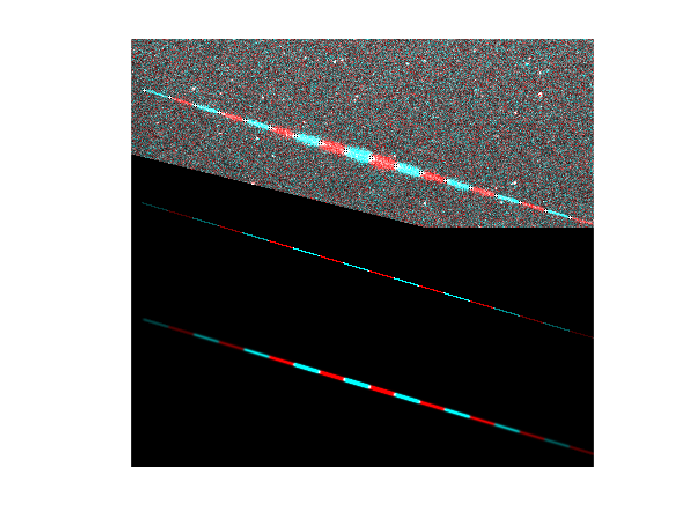}
\caption{Multi-frame meteor trace and the associated matched filter templates for thin line and PSF convolved traces (top through bottom panels respectively). The frames are sequentially displayed alternating between red and cyan.}
 \label{fig:stack}
\end{center}
\end{figure*}

However, contrary to the hybrid approach in MeteorScan, the  processing pipeline was implemented with a much faster first pass detector than the Hough transform. The pipeline starts by first identifying a potential motion candidate using a highly efficient clustering and tracking algorithm (see later section on second pass processing). The motion track then feeds the matched filter processing and refinement stage, which searches for the best fit of a propagating linear streak possessing a fixed intraframe signal level, but still allowed to vary in amplitude between frames. The thin streak model is blurred through convolution with the estimated point spread function (PSF) of the sensor system, to produce a motion template as similar as possible to the target meteor’s spatial-temporal light signature as shown in Figure \ref{fig:stack}. 

The template matching uses a maximum likelihood estimate (MLE) for its figure of merit to assess the reliability of detection. The MLE expression is given by:

\begin{equation}\label{eq2}
MLE(dB) = 5 \log_{10} \left[ \frac{ \{ \Sigma_{pixels} (S - \langle S \rangle ) R^{-1} T \}^2 }{\Sigma_{pixels} T R^{-1} T} \right],
\end{equation}

where the MLE is based on a mean removed signal S - <S> which is whitened with the inverse of the noise covariance matrix R$^{-1}$, that is then pixel-by-pixel summed over the meteor’s region of interest (ROI) by combining it with the template T over the same ROI. 

This is accumulated both spatially in the ROI and temporally across the frames possessing the meteor. The whitening is performed assuming no spatial noise correlation between pixels, which results in a diagonal covariance matrix R computed from the second order noise statistics of each pixel. This formulation follows the general matched filter formulation for moving point sources \citep[e.g][]{Mohanty1981, Shao2014, Dawson2016}.

The MLE was found to reliably identify false alarm events whose spatial-temporal pattern does not match a PSF-blurred and propagating linear streak. In addition, it was found that the matched filter template comparison to the imagery in conjunction with the MLE as a cost function, could actually be used to refine the leading-edge pick-points of the detected meteors. Previous meteor detection software determined post-detection meteor measurement positions on a per frame basis using the centroid of the streak in each frame. In fully automated processing, this could lead to spatial biasing from residual wake and/or trains that remain in the image after background subtraction. 

To avoid this effect, a better algorithm was employed that determined the leading-edge positions per frame through an iterative adjustment of the matched filter motion template (via modification of the meteor starting position and propagation dynamics), to find the best fit to the entire observed meteor trace. The advantage is that the leading and trailing edge intensity gradients (along track) in each frame, contribute more to the matched filter template fitting than the more uniform light curve of the central portions of each streak segment per frame. Thus, highly reliable leading-edge pick-points can be obtained through automated processing, by employing a non-linear minimizer, such as a particle swarm optimization (PSO) approach \citep{Eslami2012, Shi1998, Tsoulos2010}. 

The PSO method iterates on the matched filter’s MLE as the cost function metric, to converge on the best coefficient values for a meteor's dynamic motion model of position, speed, and deceleration. This was confirmed by comparing the processing pipeline's automated pick-point selections to manually measured positions of leading edges made by analysts. Note that for this to work properly in an automated mode of operation, the PSF blur must be properly estimated and incorporated into the analysis. To that end, both SExtractor and PSFEx software programs have been used to make those PSF estimates directly from the imagery.

Despite the narrow 15$^{\circ}$x15$^{\circ}$ FOV of the imaging systems in use for this project, it was found that a small percentage of meteors did not propagate across the focal plane with a constant apparent angular velocity. This turned out to be mostly due to the geometric effects of projected foreshortening rather than physical deceleration of the meteors, although the latter was a factor in some cases. The matched filter track refinement step was subsequently modified to include additional motion fitting parameters beyond just having position and velocity terms. This added deceleration and jerkiness (the time derivative of deceleration) to the motion template. 

The final formulation for the motion model is shown in Equations \ref{eq3} and \ref{eq4}. This still assumes a straight-line trajectory across the focal plane and thus the higher order moments that were added were restricted to operate strictly along the velocity vector. In the equations, X and Y are focal plane column and row positions, t is time, V$_x$ and V$_y$ are apparent angular velocities in X and Y with total speed |V|, A is the acceleration coefficient, and J the jerkiness. In general, only up to the acceleration term was found to be necessary to capture the changing apparent angle rate of all meteor encounters in our data.

\begin{equation}\label{eq3}
X(t) = X_0 + V_x(t) + \frac{1}{2}A t^2 \frac{V_x}{|V|}+\frac{1}{6} Jt^3\frac{V_x}{|V|},
\end{equation}

\begin{equation}\label{eq4}
Y(t) = Y_0 + V_y(t) + \frac{1}{2}A t^2 \frac{V_y}{|V|}+\frac{1}{6} Jt^3\frac{V_y}{|V|},
\end{equation}

As far as the authors are aware, this is the first known application of matched filter processing to optical meteor detection and measurement analysis, although the initial version of MeteorScan \citep{metscan}  in 1998 used a simple approximation to the matched filter MLE for its post-detection false alarm mitigation.

\section{IMAGE PROCESSING WORKFLOW}

\begin{figure*}
\begin{center}
\includegraphics[width=\linewidth]{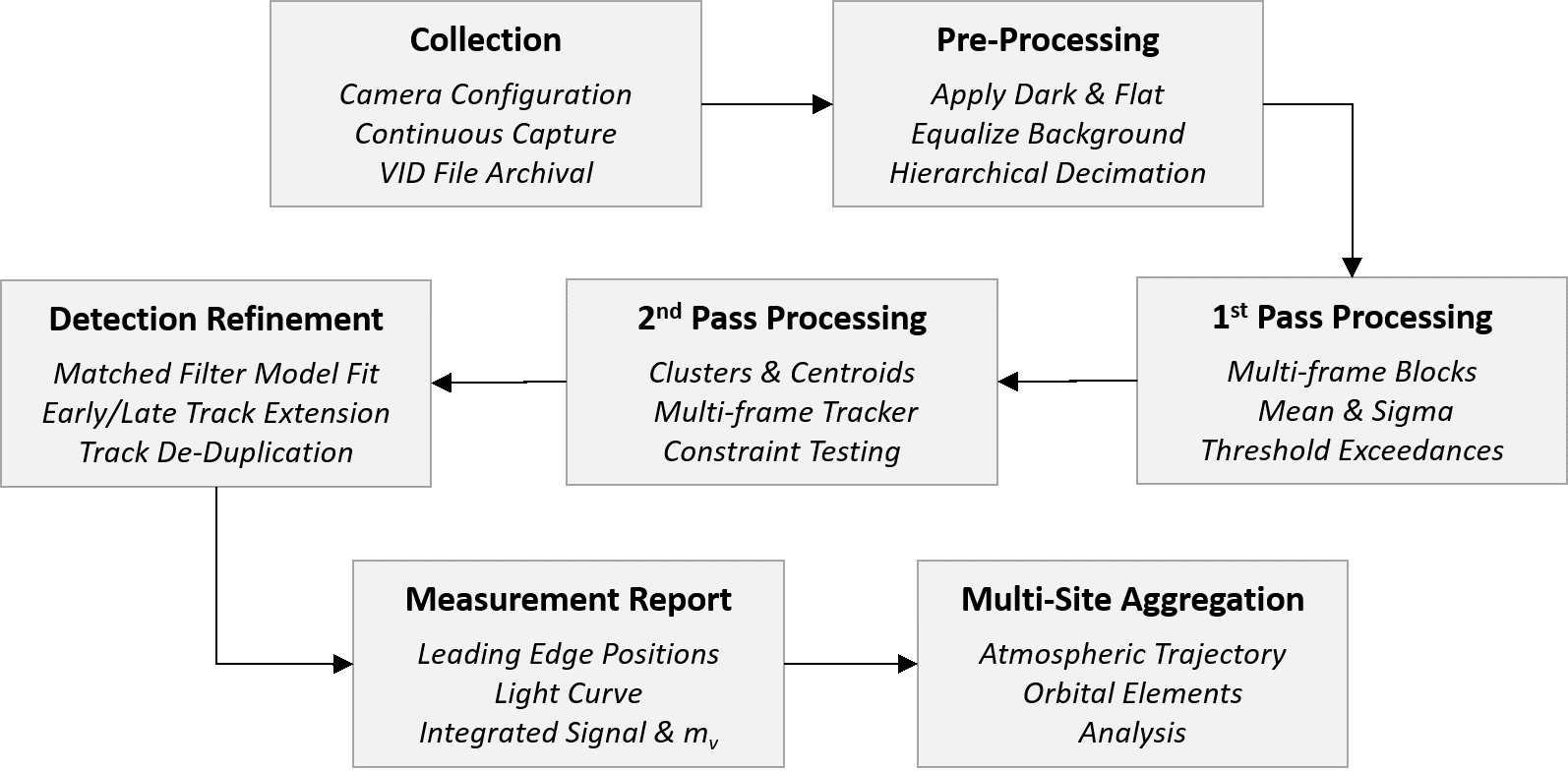}
\caption{Image processing functional work flow diagram of the EMCCD pipeline.}
 \label{fig:figworkflow}
\end{center}
\end{figure*}

The image processing components that were implemented and integrated into an end-to-end (E2E) functional pipeline involved integrating together a number of fast processing algorithms. Some were legacy capability, while others were innovations developed specifically for this project and which could be beneficial to other processing systems. Figure \ref{fig:figworkflow} shows the E2E pipeline broken down into seven major functional categories and the key processing steps. Note that one of our end goals is both meteor detection and precise, automated astrometry and photometry for each detected event on a per frame basis. 

\subsection{Collection}
The first stage in the processing chain is the collection of imagery from the EMCCD cameras. This involves configuring the camera for night operations, ensuring the camera is properly cooled, setting the region of interest, binning, using a frame rate supported by the interface bandwidth, initiating the continuous capturing of frames, and writing the raw video images to disk. 

This is executed through a series of bash scripts that use the native Nuvu SDK software together with scripts which monitor when the CAMO automated computer indicates conditions are favorable for observations. The cameras are also monitored for power, connectivity and temperature. The image resolution is based on 2x2 binning with a frame size of 512 x 512 pixels collected at 32 frames per second (fps). The imagery is stored as raw 16-bit integers in 10-minute-long file chunks with a metadata header stamp per frame. The time stamp for each camera frame is stored in that header based on an internal clock in the Nuvu EMCCD camera running at a 10 nanosecond period and tied to an embedded GPS receiver with 100 nanosecond resolution. Thus the timing is accurate to far more than our interframe period of 31 milliseconds . Note that no detection processing across the ten-minute VID file bounds is attempted at this time.

The next pipeline stage which extends from pre-processing through measurement reporting, is handled by the DetectionApplication software version 2.15 (a.k.a. DetApp) program. This was written in C++ for the highest efficiency of computation. The application is launched immediately after continuous data collection has ceased for the night, and generally wraps up before the next nightfall, thus meeting the real-time processing pace, albeit sometimes with long latency. In those rare cases when the processing does not finish before the next night's data collection starts, the file processing ceases and resumes again when observations are not being performed, until all the data are completely processed. 

\subsection{Pre-processing}
To process the stored VID file imagery for meteor detections, DetApp reads user selectable image processing control parameters, performs memory allocations based on actual image size, and ingests support data such as the latest dark file, flat file, and astrometric information. Changeable items such as the PSF parameters and the photometric zero point are accommodated by updating the control parameter configuration file on each run as necessary.

The dark file is not often updated as it was found to be very slowly changing over time. However, one dark frame is collected per night with a closed shutter, providing an archive of darks to check if the system thermal response drifts with time.  At video frame rates, the dark levels are below the sky background noise floor so it is not really critical for removal, and thus a zero dark option is provided. However, the dark file image does help to remove hot pixels. The dark is computed from a temporal median of 320 shutter closed frames.

The flat file is updated on good quality nights and is based on actual imagery using a bootstrapping approach. Because of the charged multiplication per pixel built into the EMCCD, it is not possible to collect a twilight flat without saturating the image. The method used instead involves taking a temporal average of each pixel when no star is present in that pixel, to obtain the responsivity to the background sky. Since the cameras are fixed-mounted, there is a very high likelihood that a given pixel will have a large number of frames that do not contain stellar contamination. By normalizing the per-pixel responsivity obtained to the median response level across the entire focal plane, the flat field values fall in and around unity. This approach, however, requires a previously found flat to help robustly identify star regions, and thus the flat estimation procedure is run iteratively with each flat field refinement feeding the next iteration. It was found that two to three iterations converge to a good quality flat. The flat helps to remove the variable response of each pixel element and its associated amplifier, as well as eliminate lens vignetting effects in the periphery of the image frame or dust spots associated with the lens.

The astrometric calibration file, which maps focal plane x,y cartesian coordinates to local sky's azimuth, altitude is updated every 10 minutes during normal operations. The average astrometric residuals across the field are <0.01$^{\circ}$ using a 3rd order polynomial plate fit \citep{WerykBrown2012}. 

Input frames are read into DetApp one at a time in temporal order from the VID file. Each frame first has the dark subtracted and then the resultant image is scaled by the flat field. In practice, it was found that there was a need to handle both the short-term and long-term global background level variations between frames. The former arising from distant aircraft landing lights increasing the background sky in an non-periodic fashion that the sensitive EMCCD’s could pick up as an overhead sky background level change. The longer-term temporal variations are associated with light pollution, haze level, and lunar illumination changes during the course of the night. This adjustment is an “equalization” of each image by normalizing each image to a global median background estimate (using star-free pixels only) and scaling back up to a reference background estimate. Note that to obtain a good global background estimate, the image must first be flat fielded. The global equalization helps in the robust estimation of the mean that will be used for background removal in a later step. 

The final pre-processing step is to perform a hierarchical downscaling of the imagery to reduce the effective length of longer streaked meteors. This step was required since we employed the far more computationally efficient front-end clustering/blob detector than the slower Hough transform line detection methodology. The clustering algorithm is 40 times faster computationally, but is less efficient at finding very long, thin streaks. Thus by downscaling, the streaks appear shorter in length and become more readily detectable to the clustering algorithm.

The downscaling algorithm aggregates blocks of NxN sized pixel regions in the image by taking the average of the N highest values in each region and creating a single pixel value output. The downscaled images are significantly smaller in dimension and thus processing these extra images incurs very little additional runtime cost. Once these downscaled images are run through the clustering detection algorithm, the potential meteor track is analyzed at the original full resolution of the imagery.

While our downscaling algorithm is a simple implementation, a method called perceptual downscaling \citep{Oeztireli2015}) is being explored as an alternative to the method just described to better retain sharp lines and edges at higher levels of image size reduction. 

The processing steps which are described  in the next few sections are done at several downscaled levels via parallel processing paths. The user specifies a maximum downscaling parameter Nmax where the options are 1x1, 2x2, 4x4, 8x8, and 16x16. For the EMCCD systems, only the 1x1 and 2x2 downscaled processing paths were found to be necessary, since meteors were at most 38 pixels long for the given sensor/lens configuration and the clustering algorithm works well with up to 20-pixel long streaks.

\subsection{First Pass Processing}

The processing pipeline performs multiple passes through the imagery, which requires storing in memory a ring buffer of the pre-processed images (dark removed, flattened, equalized). The number of frames held in the ring buffer is tied to a user defined maximum expected duration of a meteor event. The pipeline proceeds by working in sub-blocks of M frames, which is a number currently related to the pixel exceedance threshold. As far as detection and motion tracking of a meteor event is concerned, the transition between two temporally adjacent blocks of M frames is continuous (i.e. no meteor tracking losses occur at processing block boundaries). Each pass works through a given block of M frames before proceeding to the next block, until the end of the imagery file is reached.

The first pass through the image block is used to estimate a temporal mean $\mu$ and standard deviation $\sigma$ per pixel as well as the temporally brightest pixel exceedance map for the entire M frame block. DetApp uses the Cameras for All-Sky Meteor Surveillance (CAMS) methodology \citep{Jenniskens2011a} for generating $\mu$ and $\sigma$ associated with each block of frames, avoiding the four highest temporal values to prevent contamination by a meteor, its train, or other transient signals. This calculation is done independently per pixel across the FOV. 

The CAMS compression method for aggregating frames also finds the temporal maximum value “maxpixel” and the frame number of the maximum value “maxframe” within the block of M frames. The algorithm has a side benefit that once these two arrays are obtained, pixel exceedances are efficiently identified without direct thresholding \citep{Gural2016}. The reason for this is that the reconstructed frame sequence from the maxpixel and maxframe arrays are predominantly zero filled images with only the temporally brightest pixel excursions appearing in each frame. For the actual implementation and maximal efficiency in processing, a vector of these “exceedance” positions per frame are retained for feeding the clustering algorithm during the second pass through the block of frames.

This method of obtaining the exceedance pixels per frame has an implicit threshold set by the number of frames M used in the block. Operationally, the CAMS software worked with 256 frames, producing an effective threshold is 2.8$\sigma$ above the mean for Gaussian distributed noise. For the EMCCD system discussed herein, various values for M were tried that traded off meteor detections for false alarms (see section on detection performance). The best setting found which also had a near zero false alarm rate was M=64 frames which corresponds to an effective threshold of 2.16$\sigma$ for Gaussian noise.

Given that the $\mu$ and $\sigma$ are computed during the first pass through the block of frames, there is an alternative to finding exceedance pixels that directly thresholds each image relative to the array formed from  $\mu$+k$\sigma$ where k is a standard deviation factor. An option to do this threshold is in the processing pipeline configuration setup, but was found to produce a much higher false alarm rate than expected due to the non-Poisson distribution of the EMCCD imagery pixel values. A previously known characteristic of EMCCDs which we "re-discovered"  during our development work, is the fact that unlike most typical video meteor sensors they do not produce normally distributed noise \citep{Hirsch2013}.

\subsection{Second-pass processing}
If the first pass produces possible detections (single pixel threshold crossing exceedences), a second pass through the block of M frames is then performed and involves an attempt at finding propagating meteor segments. The pixel threshold crossers for each frame are fed into a very fast clustering algorithm \citep{Gural2016}. In this approach, a pixel exceedance is combined into a localized mini-cluster when a minimum number of adjacent pixels are also found to be exceedances. For the EMCCD, larger cells of size 32x32 then accumulate the total count of the number of these mini-clusters into these larger cell regions using a fast-indexing assignment. These cells are then further aggregated into 2x2 macro-cells and a limited number of the highest count macro-cells are then processed for centroids and associated SNR threshold detection (using the $\mu$ and $\sigma$ calculated in the first pass). While the initial clustering phase works on a binary mapping of exceedance pixels, the final clustered centroids and their associated SNRs operate on actual grey level imagery for a more robust single frame object detection.

The resulting measurement centroids for the currently processed frame are next fed to a Kalman filter, implemented as a multi-frame, multi-object tracker. The tracker uses a predictor-corrector methodology that assumes near constant velocity motion but can handle limited amounts of acceleration along track. This type of Kalman implementation is often referred to as an $\alpha$-$\beta$ tracker \citep{Blair1993} where the $\alpha$ and $\beta$ coefficients for the first seven frames ramp down from a loose constraint until more firmly fixed for the remainder of the measurement duration.

The multi-object tracker design used in DetApp has been previously described in \citet{Gural2016}. As implemented, each centroid measurement is compared to an existing set of active tracks with the goal of looking for a close association with any of their future predicted positions. An association occurs when a new measurement is within a user defined tolerance of any track’s predicted location for that time step. Note that a measurement can get associated with several different tracks, so a deduplication process is implemented down-stream once a track gets closed (i.e. after no measurements get associated to a track for a user specified number of frames).

Once a frame’s set of measurements (cluster detection centroids) have been processed through the tracker, all the tracks are updated to the current frame time either by Kalman filtering a track with a new associated measurement, or simply coasting a track via linear propagation if no measurements were associated with that track at the current time step. The status and history of each track is updated and maintained until a track is closed and passed to the detection refinement stage (see next section). Once closed, the track structure is freed up to allow for new track spawning. This process of cluster detection and tracker measurement processing continues through each of the M frames in a block in temporally increasing order.

Continuity across blocks of frames is maintained since the tracker is completely agnostic to the block processing frame boundaries.
Note also that closed tracks may have arrived at this point along different downscaled processing paths, and so are now adjusted in their measurement histories to be scaled up to the full frame resolution. All closed tracks are then tested for some basic geometric constraints to eliminate obvious false alarms. Examples are tests for linearity of the associated measurements in cross-track, spacing uniformity along-track, lower and upper apparent angular speed limits, and a minimum measurement count requirement.  All of these are made with appropriate tolerances as set by the user in the setup configuration file. Currently at least four full segments, not partially cutoff by a sensor’s FOV edge, are required for the next step of detection refinement. This allows for a reasonably good chance of producing high quality trajectories and orbits, but does bias against detection of very short duration events.

\subsection{Detection Refinement}

At this stage the separate paths for the downscaled imagery have been brought back to full resolution. Tracks that meet the minimum number of full line segment measurements are used as a motion hypothesis cue for the iterative matched filter refinement. It is at this stage where the matched filter processing is used to help further discriminate false alarms and refine the all-important leading edge pick-points along the meteor track per frame. This processing step tries to fit a propagating line segment template of varying amplitude per frame (but of fixed amplitude intraframe). This approach implicitly focuses on the leading-edge, thus avoiding the issues often encountered with centroid estimation. The computational cost for this stage is significant compared to the front-end cluster detector, but the resultant total compute load is actually very low as only a small number of clustering and tracked detections, which have passed through the earlier stages, are fed into this matched filter detection and refinement stage.

The processing proceeds by extracting an equalized imagery sequence from the ring buffer for only those frames that span the duration of the potential meteor track (where the track measurements in aggregate form a motion hypothesis). This set of images have their mean removed and are then whitened, using a diagonal covariance matrix obtained from the temporal standard deviation of pixel intensity values. This is performed on a per pixel basis, where the $\mu$ and $\sigma$ were previously obtained from the first pass block processing step. Restricting the second order noise statistics to a diagonal matrix makes the assumption that the noise is independent and identically distributed (IID) and thus there is no spatial or temporal noise correlation between pixels. The images are next convolved with the estimated PSF and thus a fast version of the matched filter template generator in the form of a thin propagating line segment, can be used for computationally efficient meteor region summation (see middle plot of Figure \ref{fig:stack}). 

The matched filter fitting (motion parameter refinement) is posed as a minimization problem where the cost function is the negative log of the maximum likelihood estimate. The log-likelihood minimum search is carried out by varying up to six motion parameters: 
\begin{enumerate}
    \item The starting position x in focal plane coordinates
    \item The starting position y in focal plane coordinates
    \item The velocity V$_x$ units of pixels/frame
    \item The velocity V$_y$ units of pixels/frame
    \item The acceleration A (optional - see Eq.\ref{eq3})
    \item The jerkiness J (optional - see Eq.\ref{eq4})
     
    \end{enumerate}

The minimum of the cost function is solved using the PSO algorithm. As currently configured, up to 5000 particles are used in the PSO to search the multi-parameter space for the best fitting motion template.

In the next step, a second more refined matched filter fit is done by generating only a mean removed and whitened image sequence, and instead convolving the PSF with the propagating line segment template. This is theoretically the same as the first matched filter fit where the PSF was convolved with the whitened imagery sequence, but in practice this second approach was found to more precisely identify leading-edge pick-points. Note however, that this second PSF convolved template is much slower to compute than the first matched filter fit with a thin line template. Since the initial fit of the motion parameters gets close to the final solution, this second phase of the matched filter fitting can be run with only 100 particles in the PSO and thus performs efficiently despite the higher computational load for each cost function evaluation.

The final motion solution to the meteor track is threshold tested against a user defined MLE constraint to further reject false alarms. Once a track is shown to have a sufficient likelihood of detection, there was found a need to extend the meteor track duration to include frames before and after the detected sequence of imagery. This was due to the discovery upon manual image inspection, of missed onset fading-in and final fading-out track segments arising from the elevated front-end threshold used in the cluster processing. The extension appends frames for meteor signal level measurements that are above an SNR = 1 and still meet the requirement of being a complete line segment per frame within the FOV. This helps to include low SNR segments missed by the initial detection screening. However, it should be noted that it is not used for further track motion parameter refinement, since the signal energy at the very begin and end of the meteor trace is usually very noisy and was found to degrade the motion parameter estimation.

\subsection{Measurement Report}

Once a potential meteor track is declared detected, attempts are made at de-duplication of the same track (or portions thereof) recorded at either the same level of downscaling from another closed track, or from detections made via other downscaling paths through the processing pipeline. This eliminates most, but not all, duplicates. A future task is to cull the remaining duplicates in a post-detection, pre-trajectory stage in the pipeline. 

One piece of reported information on a per frame basis includes the integrated signal energy less background. Note that when calculating the integrated signal levels, the equalized frame sequence spanning the meteor duration is again used, but this time a more robust temporal median estimate is estimated and subtracted per pixel, rather than removing the max excluded mean. 

The spatial integration of summed residual pixels “P” is performed over a region-of-interest (ROI) defined by the refined motion track in a given frame and spread by the PSF. The PSF extent is broadened further by the estimated signal energy in the streak for each frame, which accounts for the streak’s increased cross-track extent in brighter portions of the meteor light curve. This ROI is further enhanced by a factor of two to ensure that all signal energy in the tails of the PSF is included in the integrated pixel sum. The meteor magnitude in a given frame is then calculated from the log sum pixel value and scaled by the zero-point magnitude of the camera as

\begin{equation}\label{eq5}
M(frame\,k) = M_{zero-point}+2.5 \log_{10} \Sigma_{ROI(k)}P
\end{equation}

The zero-point magnitude is computed on a per event basis using all available stars in the field of view and thus determines the zero-point intercept through an automated photometric calibration. The zero-point was found to drift very little both during the night (<0.1 mag) and remain stable from night to night. 

The final report generation is done in a custom “ev” (event) text file format for each detected track. Included in the report files are the UT time, motion parameters, and leading-edge position measurements in focal plane coordinates, the corresponding local azimuth, altitude, right ascension and declination referenced to J2000, log sum signal level, and magnitude.

\subsection{Multi-Site Aggregation }

Once the DetApp software has completed the processing of the video files for all the cameras from a given night at a given site, the detected tracks are uploaded to central servers. The next stage of processing involves the multi-site aggregation of tracks. This step searches for spatially and temporally correlated events from multi-site cameras to form two-station triangulation and atmospheric trajectory estimation. This is done using long standing infrastructure and an analysis code base as part of the ASGARD software suite \citep{Weryk2008}. It works fundamentally by finding temporal coincidences and linking those common events across sites that fall within a one second time window. 

Space-time aggregated tracks are fed into several trajectory estimation codes to allow for the comparison and verification of processing results. These include MILIG and SmeTs which are similar implementations of a least-squares fit to straight line trajectories \citep{Borovicka1990} as well as a multi-parameter fit (MPF) trajectory routine \citep{Gural2012}. Of note is that the MPF algorithm has been enhanced for this project, to make the algorithm performance more robust and easier to interface. At its core, the MPF algorithm attempts to solve for the trajectory unknowns of the state variables in position and velocity, motion dynamical parameters, and relative camera timing offsets, all fit simultaneously using a cost function minimization approach. 

The enhancements to MPF that were made as part of this project include the introduction of a single I/O structure, software function modularization, and replacing the simplex amoeba algorithm with the PSO for better global minimization \citep{Egal2017}. Further improvements included the addition of new angle formats for the input measurements (all converted to ECI for internal computations) and a new cost function that is weighted by the user input of measurement error on a per measurement basis rather than per camera track (to account for light curve variability that results in measurement error variability along the track from the changing SNR).

In addition, a new velocity estimator was used in the boot-strapping phase that was not based on positional differences over time, but a least mean squares (LMS) solution for a constant velocity model to the positional measurements, thus avoiding the root-two differencing increase in noise \citep{Gural2018}.  The MPF algorithm proceeds through a boot-strapping operation, starting with an intersecting planes solution \citep{Davidson1936, Ceplecha1987} that feeds a LMS parameter estimate of the state variables \citep{Borovicka1990} and finally produces a simultaneous MPF solution of all the unknown trajectory parameters. Output products from the MPF include the trajectory parameters and their Monte Carlo based noise estimates, along with per frame measured and modeled estimates of meteor 3D position and velocity. The trajectory is then fed into the orbital determination routine described in \citet{vida2020a} 

\section{Detection Performance}

Several trade studies and simulation tests were done to optimally select run time configuration settings and fine tune the meteor detection performance for the EMCCD cameras. There are a number of parameters to adjust for detection thresholds which impact the best operating performance for a given sensor’s characteristics. The runtime configuration parameter file listing as finally adopted for the EMCCD cameras is shown in Appendix B. While this emphasizes the large number of settings that need to be adopted for a given instrument setup, in practice only a few need to be explored to optimize the performance of the detection phase.

\subsection{Parameter Trade Study on Real Imagery}
A study was done varying just a few of the critical parameters which were known to most influence the probabilities of detection and false alarms. This was accomplished by running DetApp against actual collected meteor imagery through a range of the minimum required nearest neighbor exceedance count to declare a mini-cluster, the MLE threshold for matched filter detection, and the number of frames M used in a processing block. The latter changes the effective front-end threshold for pixel exceedances. 

\begin{figure*}
\begin{center}
\includegraphics[width=\linewidth]{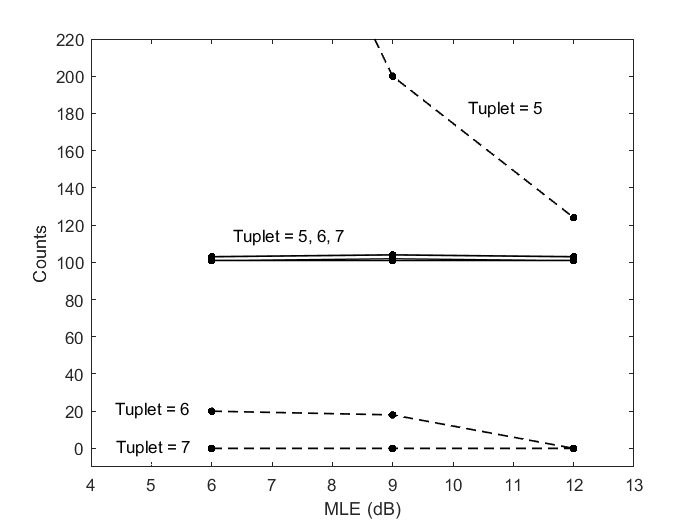}
\caption{Counts of true positives (solid curves) and false positives (dashed curves) for M = 32 frames per block.}
 \label{fig:truefalse}
\end{center}
\end{figure*}

The cluster neighbor “tuplet” count controls how many threshold exceedance pixels must be nearest neighbors to any given threshold crossing pixel for a detection. Meeting or exceeding this count identifies a mini-cluster of pixels. If too few nearest neighbors are used for the threshold, the numbers of clusters would grow rapidly with false alarm rates rising dramatically. At the opposite extreme, too high a neighbor count requirement and faint meteors that have very few neighboring exceedance pixels will not get clustered together and thus not tracked into a multi-frame detection. The optimal count for a reasonably low false alarm rate was initially believed to be a tuplet value of 6 as seen in Figure \ref{fig:truefalse}. 

The detection performance shown by the solid lines was found to be insensitive to other parameter variations. But the false alarm rate grew as the clustering tuplet size decreased, as expected. However, this was first studied at a block frame count size of 32 frames and more optimal results were found when running 64 frames with a tuplet size of 5. Thus, the current operational configuration requires at least five exceedance neighbors to form any given mini-cluster.

The second optimization involved choosing the best MLE threshold as used by the matched filter refinement stage. This can be used for false alarm mitigation by setting a higher MLE threshold, but the trade-off is an increased number of missed faint meteors. In general, this threshold was found to not influence the detection performance and simply helped limit the false alarm rate. Using a MLE=0.0 dB cutoff was found to be adequate since this threshold was applied so late in the processing pipeline, that most false positives were already removed. 

From experimentation, it was found that the third and most critical parameter for optimal detection is the number of frames M aggregated for the mean, sigma, maxpixel, and maxframe arrays. The latter two array products control which pixels get declared as an exceedance threshold crosser. The fewer the number of frames used in the aggregation, the lower the effective threshold. This results in a higher number of exceedance pixels and clusters, and ultimately higher false alarm counts. Using too many frames with a much higher effective threshold, may result in many faint meteors near the noise limit of the imagery being missed. 

\begin{figure*}
\begin{center}
\includegraphics[width=\linewidth]{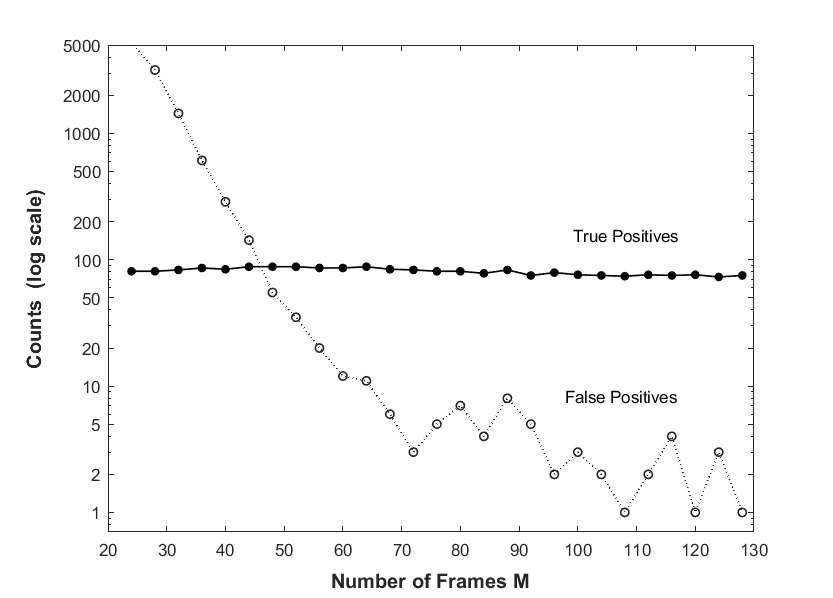}
\caption{Counts of true positives (solid curves) and false positives (dashed curves) as a function of the Maximum Likelihood Estimate (MLE) for various numbers of frames per block. This is equivalent to the effective pixel exceedance threshold.}
 \label{fig:exceedance}
\end{center}
\end{figure*}

An optimal range for this value was found by having an analyst examine actual meteor video of the software pipeline’s detections, with the best performing result falling in a broad range from 32 to 64 frames. Figure \ref{fig:exceedance} shows that the true positive detection performance is fairly insensitive to frame count, but the false alarm rate did begin to rise as the exceedance pixel density increases with the lower range of frame count M (i.e. lower threshold).  On this basis, it was decided to operate with nearly zero false alarms at a block frame count of 64.  

\subsection{Simulation Based Algorithmic Enhancements }

During the study which examined actual imagery involving visual inspection by a trained analyst, it was found that there were some missed meteors that should have been easily detected by the software. Thus, to further fine tune the detection algorithm, a moderate fidelity simulation was implemented in Matlab to allow rapid prototyping, implementation, and test case turnaround. This had an advantage over using actual imagery since it provided a well-controlled environment with ground-truth values for various meteor feature parameters, but with sufficient fidelity to characterize difficult detection cases. The approach used was to generate synthetic VID files from a Matlab simulator that could then be ingested and processed by the C++ DetApp executable.

The simulation generated meteor sequences by  producing a randomly distributed noise background that was added to a fixed bias level for all frames in the sequence. One meteor of duration ranging from 3 to 30 frames long was injected per file, with the meteor starting in frame 32 of a 256-frame sequence. The meteor amplitude (brightness) was set to be either temporally constant, to have a parabolic light curve that varied in magnitude with time \citep{Campbell1999} or could be chosen as a late peaking light curve mimicking single body ablation \citep{Ceplecha1998e}. 

\begin{figure*}[h!]
\begin{center}
\includegraphics[width=\linewidth]{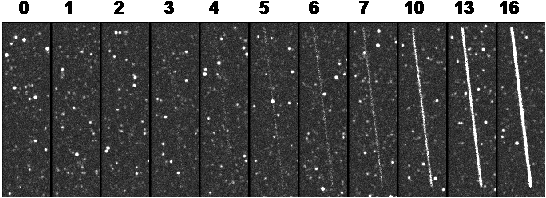}
\caption{Examples of equivalent stacked frames for simulated meteors with peak SNRs ranging from 0 to 16 dB (left to right) for a constant light curve.}
 \label{fig:constantlc}
\end{center}
\end{figure*}

A fixed-width Gaussian PSF was applied to the meteor trace to spread the streak in both along-track and cross-track directions (see Figure \ref{fig:constantlc}). Note that when noise was added, the standard deviation used in the Gaussian random draw was based on a power law of the mean pixel level (i.e., background plus meteor, done independently per pixel) with an exponent of one-half to represent Poisson type noise statistics. This was the closest representative noise distribution to the actual EMCCD imagery characteristics, where several exponent powers were also tried that had little influence on the performance results.

Meteors were assumed to have constant apparent angular velocity across the focal plane and covered the full range of speeds allowable on dynamical grounds (12.3-71.9 km/s). Given the per pixel angular resolution and frame time spacing of the EMCCD systems, this simulated the full range of focal plane angular velocities encountered in real imagery. The frames were built to be oversized with meteor starting positions and directions selected randomly both inside and outside the FOV, such that the frames were then trimmed to the actual EMCCD processing dimensions. This was to properly account for meteors entering, exiting, or corner cutting the FOV with partial length streaks at the edges. 

\begin{figure*}
\begin{center}
\includegraphics[width=\linewidth]{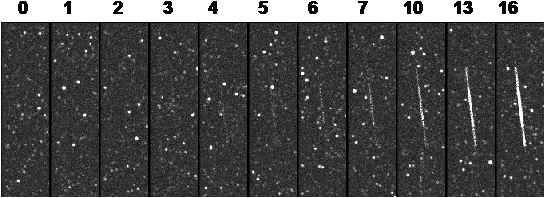}
\caption{Examples of equivalent stacked frames for simulated meteors with peak SNRs ranging from 0 to 16 dB (left to right) for a parabolic light curve.}
 \label{fig:paraboliclc}
\end{center}
\end{figure*}

It should be noted that no attempt was made to use a true magnitude distribution or velocity distribution that represented the real meteor environment encountered by the Earth. Rather, the simulation generated a full range of realizable meteor magnitudes and speeds which aimed to uniformly sample all possible encounter values and present the detection algorithm with as wide a range of signal levels and streak dynamics as possible. This helped to pinpoint weaknesses in the algorithm and fine tune the software implementation.

The initial results using the simulation clearly identified a number of cases that were missed by the detection algorithm, yet had visually discernable and detectable meteor traces. These were identified mostly as “edge” cases where there were:
\begin{itemize}
    \item Meteors crossing the boundary of the FOV with one frame having a cutoff segment length
    \item Meteors that ran close to and parallel to the edge of the FOV
    \item Meteors that were culled too aggressively after their tracks were extended
\end{itemize}

Simple modifications to the processing pipeline addressed these issues and resulted in dramatically improved detection reliability. The changes implemented increased the number of detections on actual imagery by nearly a factor of two. This latter effect was mostly due to the higher-than-expected number of meteors that occur along an edge or cutting through the corners of the FOV. By changing the culling criteria to test that both the begin and end points of a modeled track segment per frame were within the FOV, more meteors along the edge were accepted and partial segment tracks were avoided, thus not failing the “uniformity of measurement spacing” criteria.

\begin{figure*}
\begin{center}
\includegraphics[width=\linewidth]{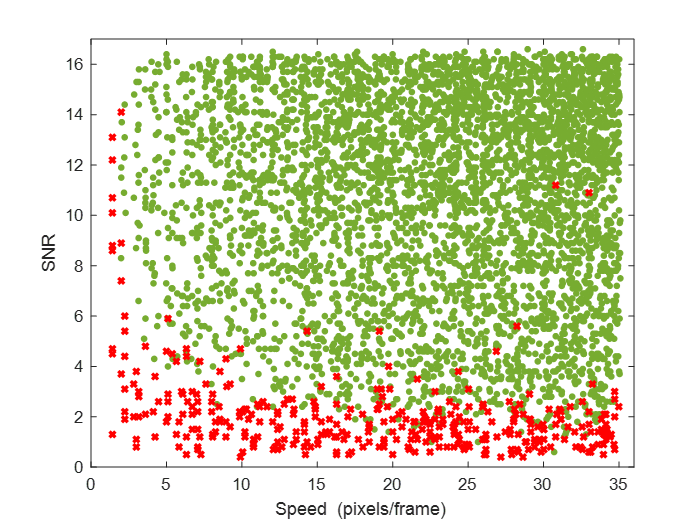}
\caption{SNR versus speed in pixels per frame for synthetic meteors having constant intensity light curves. Green Dot = detection. Red X = missed detection}
 \label{fig:snrspeed}
\end{center}
\end{figure*}

After these modifications were implemented, a set of simulation runs were performed to evaluate the detection efficacy of the algorithm as a function of both SNR and meteor speed on the focal plane. The results are shown in Figure \ref{fig:snrspeed}. The choice of either late peaking light curves or parabolic light curves did not noticeably affect the results so only the parabolic is shown for the second set of figures. 

\begin{figure*}
\begin{center}
\includegraphics[width=15pc, height=22pc]{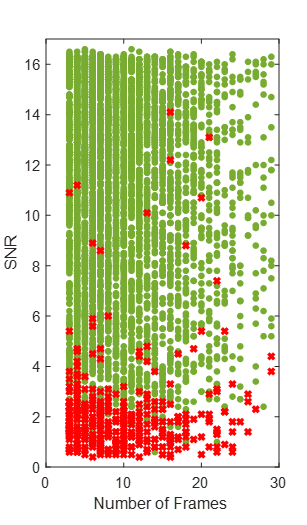}
\caption{SNR versus number of frames for meteors having constant intensity light curves. Green Dot = detection. Red X = missed detection}
 \label{fig:snrdur}
\end{center}
\end{figure*}

Figure \ref{fig:snrspeed} shows the SNR versus apparent angular speed for detections (solid green dots) and missed detections (red x’s), while Fig \ref{fig:snrdur} is SNR versus duration of the meteor in number of frames. It is clear that the detector misses the lowest SNR cases as expected. However, they are so weak in intensity that the meteors are essentially buried in the noise and barely discernable (see Figure \ref{fig:constantlc} to see a visualization of meteor images for various SNR levels).

A higher SNR cutoff occurs for more realistic meteors with a rising and falling light curve (parabolic magnitude in time). This is because early and late portions of the meteor track are close to the noise floor and if those end frames are noise buried on a low frame count (short duration) meteor, it is more likely not to get detected given the four full-frame requirement. Some high SNR cases were also missed, but these were typically cases with an extremely low apparent velocity of less than 2 pixels per frame; which looked essentially like stationary objects to the detection algorithm. In general, the detection algorithm performance was found to be insensitive to speed (effectively the meteor line segment length per frame) as well as being insensitive to the number of frames in duration, indicating that the probability of detection was primarily driven by SNR. 

\begin{figure*}
\begin{center}
\includegraphics[width=\linewidth]{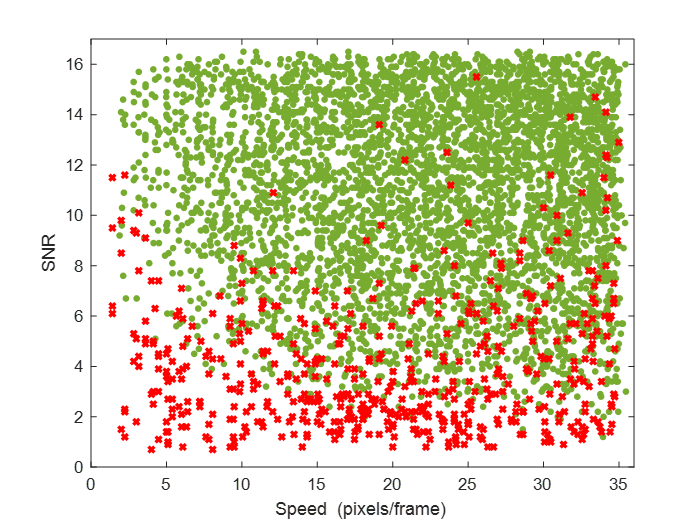}
\caption{SNR versus speed in pixels per frame for synthetic meteors having parabolic magnitude light curves. Green Dot = detection. Red X = missed detection.}
 \label{fig:snrspeed-parab}
\end{center}
\end{figure*}

\begin{figure*}
\begin{center}
\includegraphics[width=15pc, height=22pc]{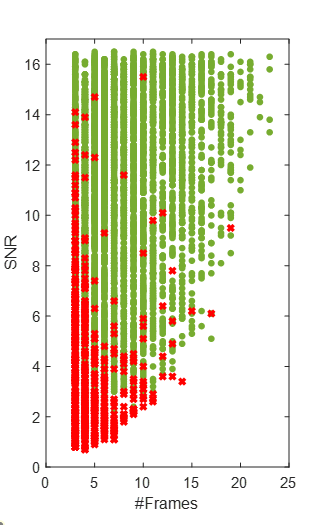}
\caption{SNR versus speed in pixels per frame for synthetic meteors having parabolic magnitude light curves on the left panel, and corresponding SNR versus frame duration on the right panel. Green Dot = detection. Red X = missed detection.}
 \label{snrdur-parab}
\end{center}
\end{figure*}

\begin{figure*}
\begin{center}
\includegraphics[width=\linewidth]{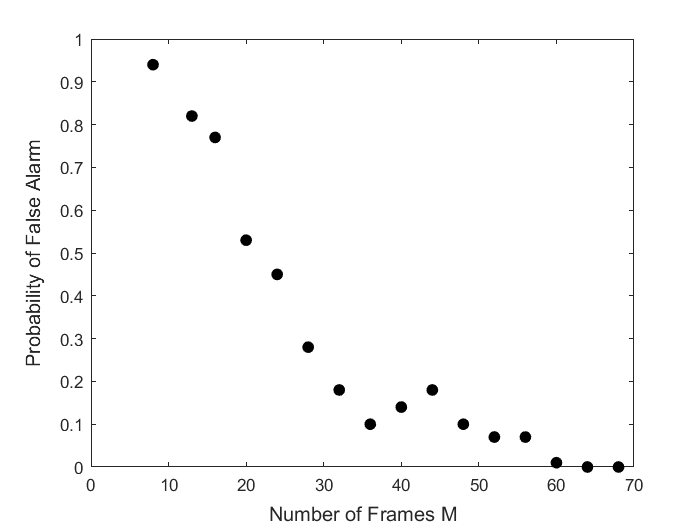}
\caption{Probability of false alarms per 100 random simulation files as a function of number of frames M in a processing block. Effective thresholds of 1.7$\sigma$ 1.8$\sigma$ 1.9$\sigma$ 2.0$\sigma$ 2.1$\sigma$ and 2.2$\sigma$ correspond to M=23, 28, 35, 44, 56 and 72 frames respectively. There are essentially no false alarms for both simulation and actual imagery at or above M = 64.}
 \label{fig:probfalse}
\end{center}
\end{figure*}

During this evaluation, the false alarms rates in both the simulation runs and on actual imagery were found to be negligible for the runtime parameter settings used in operations on real EMCCD imagery. To test whether the code does produce false alarms on simulated imagery, the number of frames per block processed M was significantly lowered. This is equivalent to lowering the front-end detection threshold for producing exceedance pixels. Figure \ref{fig:probfalse} shows the impact of changing the frame count M per block when processing simulated meteors. This validates the decision to use 64 frames in operations, resulting in extremely few false alarms.

\subsection{Full Matched Filter Detection Comparison}
To compare our hybrid processing approach to one where a full matched filter algorithm is used as the primary front-end meteor detector, six separate ten-minute collection VID files from a clear night on August 26, 2020 were singled out for detailed analysis. These archived files were processed with the current operational settings of the detection application, using a code modification that replaced the cluster detector with a matched filter detector. A different handling of the imagery sequence needs to be done as the matched filter tries to take advantage of the SNR gain from multiple frames aggregated together before threshold detection. A sliding window of four frames was used to obtain the multi-frame integration gain for the matched filter, with the intention that duplicate (spatially shifted) tracks from adjacent four frame sequences could be stitched back together in a post-detection review process. The goal was simply to see how much fainter the matched filter detector would perform on real data (which is theoretically the best detector possible in Gaussian noise) , and not try to design an optimal way of processing imagery with a front-end matched filter. The latter would need to be explored if a GPU based implementation were to be pursued.

For each four-frame imagery sequence processed, a quadruple nested loop was implemented covering the possible starting positions X and Y as well as speeds Vx and Vy, which generated an MLE estimate for each unique motion hypothesis. Any FOV edge crossing hypotheses and speeds outside a realizable minimum and maximum velocity constraint, were culled from the motion hypothesis list. This left approximately 0.8 billion hypotheses to be evaluated per four-frame set, with only the highest MLE amongst them to be reported.

On a single processing core this would typically take 0.8 hours of CPU wall clock time for a four-frame set. The frame set was then shifted forward in time by one frame, and this process repeated 19,200 times to cover all the frames in a VID file. Distributing this process on a cluster across the 48 cores of a single node, took 14 days to complete the processing of one ten-minute VID file. 

After an initial analysis, it was found that culling the reported list of results at a MLE threshold of 6.0 dB and requiring object motions of greater than 2.25 pixels per frame, resulted in good thresholds for detecting some of the faintest meteors in the imagery, right down to the imagery’s noise level (see Figure 11 for the rising false alarm counts as a function of lowered MLE level). 

The list of candidate detections from the matched filter runs were each visually inspected by hand using a video playback loop and then classified by an analyst as either false alarm (noise, hot pixel flares, satellite glints) or as a meteor if visually discernable above the noise background. In each case where a DetApp detection of a meteor occurred with the pipiline's hybrid algorithm, there was a corresponding front-end matched filter detection as well, but not vice versa when extremely faint meteors were involved. This is shown in Figure \ref{fig:matched-filter} where the matched filter ONLY detections appear as an excess above the jointly detected MF and DetApp meteor counts. In total, nearly three times as many meteors were detected with the matched filter than our approach, but at a very high cost (approximately four orders of magnitude) of added computational load.

\begin{figure*}
\begin{center}
\includegraphics[width=\linewidth]{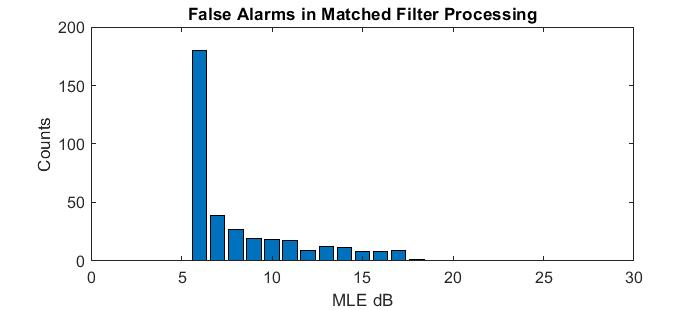}
\caption{Number of meteor false alarm detections as a function of chosen MLE for the matched filter.}
 \label{fig:MLE-false}
\end{center}
\end{figure*}

The performance difference in the matched filter versus DetApp MLE metric is that the full matched filter is 3 dB more sensitive in detection. This corresponds to about a 3/4 meteor magnitude limit improvement in sensitivity. However, the visual inspection of the very faintest of those meteors, indicated that an analyst would be hard pressed to provide reliable leading edge pick points, let alone have an automated system perform those measurements accurately. 

In this sense applying a full MF approach might be useful in limited scenarios where the intent was to estimate populations of faint meteors (flux counting) but not determine trajectories and orbits.

It may still be possible to push down the overall sensitivity of th pipeline by only 2 dB, or roughly half a magnitude, and nearly double the number of meteors with reliable measurements. This is a research task in progress to find a more sensitive front-end detection algorithm that still meets the next day processing constraint.

\begin{figure*}
\begin{center}
\includegraphics[width=\linewidth]{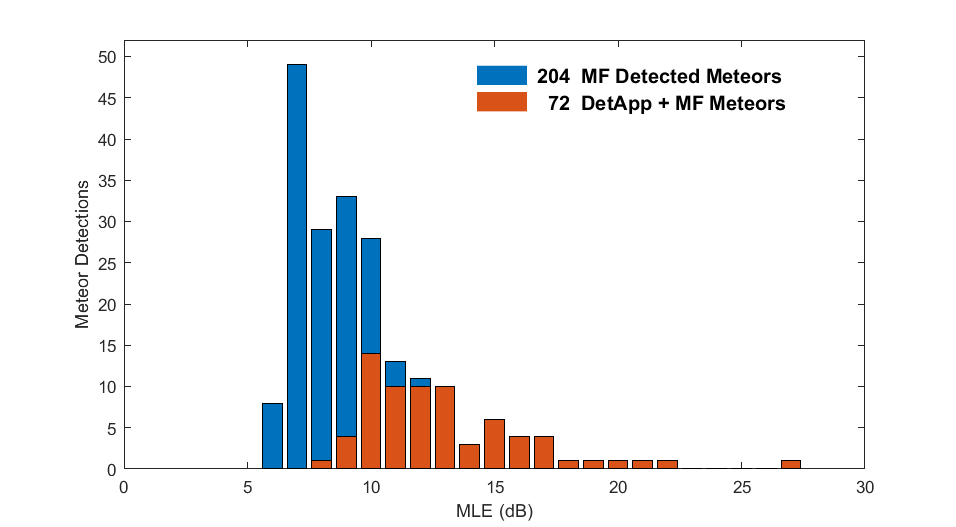}
\caption{Numbers of meteors detected by only the matched filter algorithm (blue) and jointly by both the DetApp and matched filter applications (red), plotted versus the MLE metric. Note the increased sensitivity of the matched filter meteor detection at lower MLEs (4 dB MLE change = 1 magnitude).}
 \label{fig:matched-filter}
\end{center}
\end{figure*}

\subsection{Preliminary Results in Processing Actual Imagery}

Through daily operations covering the period from February 2017 to December 2020, the EMCCD camera systems have been collecting multi-site imagery, performing meteor detection, and computing atmospheric triangulation and orbits. The yield has been better than expected with slightly more than one meteor per minute per camera. 

\begin{figure*}
\begin{center}
\includegraphics[width=\linewidth]{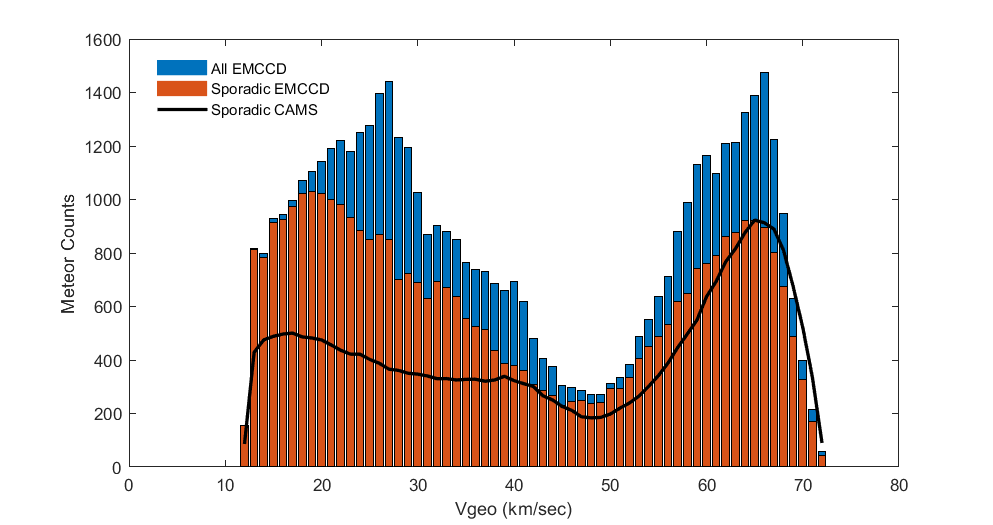}
\caption{Plot of the geocentric velocity distribution for all meteors and just sporadics as seen by the EMCCD system compared to the CAMS sporadics (CAMS 2016) which have been normalized to the EMCCD sporadic peak at 66 km/sec, showing significant numbers of low-velocity smaller-mass meteors detected by the EMCCD system.}
 \label{fig:velocity-dist}
\end{center}
\end{figure*}

Thus far the system has collected 54642 triangulated meteors. While this paper does not try to de-bias or draw conclusions from these initial raw results, they are presented here to show the potential contributions this system has to offer in future detailed analysis and meteor modeling. It is apparent that the system does capture a large fraction of slow velocity sporadic meteors as seen in Figure \ref{fig:velocity-dist}, which is a similar observation seen with the previous generation low-light intensified imaging systems that have been in operation for almost a decade as part of the Canadian Automated Meteor Observatory \citep{Campbell-Brown2015}. 

\begin{figure*}
\begin{center}
\includegraphics[width=\linewidth]{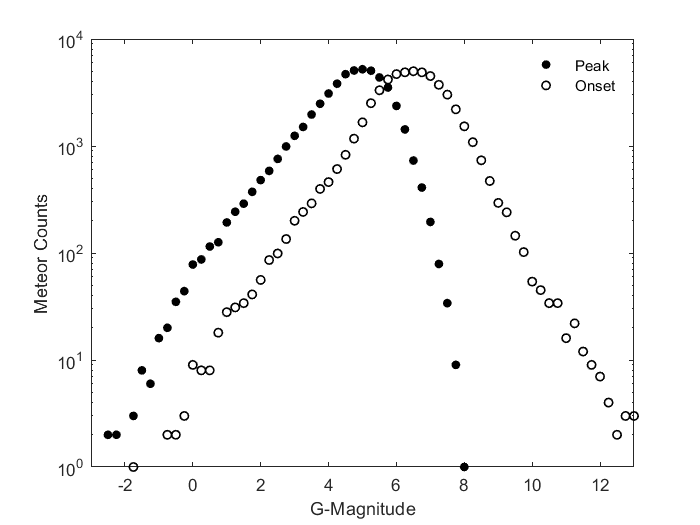}
\caption{Plot of the onset and peak magnitude distribution as measured by the DetApp pipeline on detected meteor EMCCD pairs showing completeness to peak magnitude +4.5 with the typical meteors observed around +6.5.}
 \label{fig:fig-mag}
\end{center}
\end{figure*}

To separate out shower meteors from sporadics, a radiant position and velocity association is performed as described in \citet{Vida2021}, such that a shower meteor must be within a 3 degree radiant distance and have relative velocity within 10 percent, when performing a search through a shower table. Otherwise the meteor is declared a sporadic. The percentage of sporadic orbits was found to be 76\%, a value for the smaller mass meteoroids (magnitude roughly +5)  similar to that  measured with the co-located Canadian Meteor Orbit Radar \citep{Brown2008}, which finds 90\% of all meteoroids to be sporadic at a limiting magnitude of +7.  From the magnitude distribution in Figure \ref{fig:fig-mag}, the typical meteors seen by this system is near +6.5 meeting the desired goals of the project. The magnitude histogram also shows a completeness down to M$_G$ ~ +4.5 where the slope implies a magnitude distribution index for sporadics of 2.67. This corresponds to a differential mass distribution index of s = 2.07, comparable with other measurements of the sporadic background at these magnitudes \citep{Pokorny2016}. 

\begin{figure*}
\begin{center}
\includegraphics[width=\linewidth]{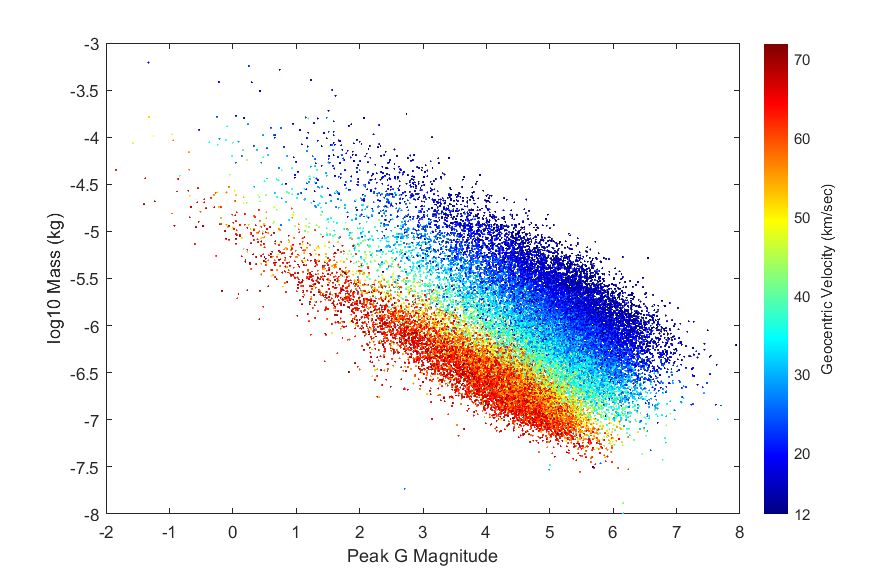}
\caption{Plot of the log mass versus magnitude for sporadic meteors showing the velocity dependence and the system’s sensitivity to low velocity meteors down to mass 3x10$^{-7}$ kg.}
 \label{fig:mag-dist}
\end{center}
\end{figure*}

On particularly transparent nights, several meteors were detected with peak magnitudes as low as +8, with onset and fading near +10 magnitude. Figure \ref{fig:mag-dist} shows the mass versus magnitude relationship with clear segmentation by geocentric velocity. Here the mass is computed assuming a constant luminous efficiency of 0.7\% based on \citet{Subasinghe2018} which suggests this is a good average for our median mass meteoroids of 10$^{-6}$ kg. We then use a standard luminous power of 945W for a 0 mag meteor in the G passband \citep{Brown2017} to compute total energy by integrating under the light curve. The system is reliably collecting meteors below 20 km/sec for masses as small as 3x10$^{-7}$ kg. More detailed analysis are in \citet{Mills2021} and in a forthcoming paper (Mazur et al. 2022). 

\section{Summary/Conclusions}

We have described the design, implementation and performance evaluation of a robust software processing pipeline to execute a multi-station survey of the faint meteor population. This has involved the deployment of highly sensitive EMCCD cameras coupled together with the automated collection and processing of meteors near the noise limit of the sensors. By working in a magnitude range centered on +6.5, this system can coincidentally collect the optical counterparts to the underdense radar echoes as seen by CMOR and the high-resolution imagery collected by the mirror-tracking system of CAMO \citep{Weryk2013, Vida2021}. 

The system sees a larger relative population of low velocity sporadic meteors compared to other video systems. While the origin of this difference remains unclear, whether it be past imaging system biases or a real enhanced low velocity distribution due to dynamical evolution of smaller mass meteoroids, the imaging system should help characterize the sporadic population’s physical and orbital characteristics. The data rates are sufficiently high to help measure flux, point of origin and mass distribution.

The achievement of these goals was enabled through the careful and innovative development of an automated image processing suite that employs a matched filter for both false alarm mitigation and the precise refinement of leading-edge measurements per video frame. A fast front end detector permitted the use of the matched filter within the pipeline without requiring expensive computational capability. Some innovations in processing that may be applicable to other meteor processing systems include imagery-based flat fielding, hierarchical downscaling, the use of fast clustering vice line detection algorithms, and simulation-based algorithm performance optimization. A key discovery with the latter is that there can be significant biases when sampling the meteoroid population given the limited size of a sensor’s FOV. Besides the many innovations outlined in this paper, the system also uses well-established image processing, astrometric and photometric techniques as commonly used in the meteor community.

Our central result has been a first time demonstration of the operational benefit of matched filter processing to detect the faintest meteors possible in an optical system, while also yielding high quality metric data of focal plane positional measurements in a fully automated setting. Future papers will explore the analysis of the unique data set this collection system brings to meteor science.

\bmhead{Acknowledgements}
The authors thank Dr. William Cooke, lead of NASA’s Meteoroid Environment Office, for technical discussions and comments. Funding for this work was provided in part through NASA co-operative agreement 80NSSC21M0073. This work was funded in part by the Natural Sciences and Engineering Research Council of Canada Discovery Grants program (Grants no. RGPIN-2016-04433) and the Canada Research Chairs program. 

\section*{Declarations}

Author contributions:

All authors contributed to the original equipment design, ongoing data collection operations and daily analysis of measurements. Material preparation and analysis for this paper were performed by Peter S. Gural with much of the bulk post-detection processing done by Tristan Mills. The first draft of the manuscript was written by Peter S. Gural, with the science applicability reviewed and edited by Peter B. Brown. All authors commented on previous versions of the manuscript and have read and approved the final manuscript.

Compliance with Ethical Standards:

The authors have no conflicts of interest to declare that are relevant to the content of this article. There was no reasearch that involved humans or animals. Funding sources have been disclosed in the Acknowledgements section. All authors certify that they have no affiliations with or involvement in any organization or entity with any financial interest or non-financial interest in the subject matter or materials discussed in this manuscript. The authors have no financial or proprietary interests in any material discussed in this article.The authors have no relevant non-financial interests to disclose.

\begin{appendices}

\section{Appendix A: Summary of Algorithm Methodologies}\label{appendix-a}

A comparison of potential algorithms that could be used
at various stages of the processing pipeline are summarized in Table
A-1. This is by no means an exhaustive review of possible processing
algorithms, but is fairly representative of methods used in a variety of
legacy meteor detection applications developed over the past two decades
and applied to video imagery.

Table A-1. The pros and cons of various processing pipeline algorithms

\clearpage
\onecolumn

\begin{landscape}

\begin{longtable}[]{@{}lll@{}}
\toprule
\textbf{Algorithm} & \textbf{Advantages} & \textbf{Disadvantages}\tabularnewline
\midrule
\endhead
\multicolumn{3}{c}{\textbf{Pre-Processing -} \emph{Clean-up of each frame's raw artifacts}}\tabularnewline
\midrule
\begin{minipage}[t]{0.30\columnwidth}\raggedright
Dark Removal\strut
\end{minipage} & \begin{minipage}[t]{0.30\columnwidth}\raggedright
Generally, not needed at video rates.

Eliminates or masks sticky bit pixels.

Long term stability = infrequent updates.\strut
\end{minipage} & \begin{minipage}[t]{0.30\columnwidth}\raggedright
Need to take a cover closed collection.\strut
\end{minipage}\vspace{6pt}\tabularnewline
\begin{minipage}[t]{0.30\columnwidth}\raggedright
Flat Fielding\strut
\end{minipage} & \begin{minipage}[t]{0.30\columnwidth}\raggedright
Removes pixel-to-pixel sensitivity variances.

Can use a good quality shifting sky collect.

Long term stability = infrequent updates\strut
\end{minipage} & \begin{minipage}[t]{0.30\columnwidth}\raggedright
Twilight flats - difficult to obtain uniformity.

Using night imagery - requires star ID.\strut
\end{minipage}\vspace{6pt}\tabularnewline
\begin{minipage}[t]{0.30\columnwidth}\raggedright
Hot Pixel Removal\strut
\end{minipage} & \begin{minipage}[t]{0.30\columnwidth}\raggedright
Eliminates flaring pixels and cosmic rays.

Usually not confused with meteors.\strut
\end{minipage} & \begin{minipage}[t]{0.30\columnwidth}\raggedright
Extra processing load on every frame.\strut
\end{minipage}\tabularnewline
\midrule
\multicolumn{3}{c}{\textbf{Clutter Suppression -} \emph{Stationary Object Removal and Noise Mitigation}}\tabularnewline
\midrule
\begin{minipage}[t]{0.30\columnwidth}\raggedright
Frame Differencing\strut
\end{minipage} & \begin{minipage}[t]{0.30\columnwidth}\raggedright
No need for dark removal or flattening.

Removes fixed pattern noise and stars.

Removes slow background variations.

2x faster if absolute value of difference.

Star drift not significant at video rates.

No wake/train if signed frame difference.\strut
\end{minipage} & \begin{minipage}[t]{0.30\columnwidth}\raggedright
Introduces square root of two more noise.

Must difference back many frames for allsky.

Wake/train remains if abs difference.\strut
\end{minipage}\vspace{6pt}\tabularnewline
\begin{minipage}[t]{0.30\columnwidth}\raggedright
Mean Subtraction\strut
\end{minipage} & \begin{minipage}[t]{0.30\columnwidth}\raggedright
Removes fixed pattern noise and stars.

Square root of two noise NOT introduced.

Median preferred over mean -- but costly.\strut
\end{minipage} & \begin{minipage}[t]{0.30\columnwidth}\raggedright
Requires an efficient mean update.

Star drift a problem in very narrow FOV.

Recommend flat fielding of background.

Equalize any global background variation.

Wake/train not removed = impacts centroids.\strut
\end{minipage}\vspace{6pt}\tabularnewline
Whitening & Suppresses high variance pixels (e.g. stars). & Requires
estimation of 2\textsuperscript{nd} order noise stats\tabularnewline
\midrule
\multicolumn{3}{c}{\textbf{Thresholding -} \emph{Identification of pixel exceedances above the mean}}\tabularnewline
\midrule
\begin{minipage}[t]{0.30\columnwidth}\raggedright
Fixed level above the mean\strut
\end{minipage} & \begin{minipage}[t]{0.30\columnwidth}\raggedright
Only mean needs to be tracked or\ldots{}

frame difference assumes zero mean.\strut
\end{minipage} & \begin{minipage}[t]{0.30\columnwidth}\raggedright
Doesn't handle different variances per pixel.\strut
\end{minipage}\vspace{6pt}\tabularnewline
\begin{minipage}[t]{0.30\columnwidth}\raggedright
Mean + factor * root-mean\strut
\end{minipage} & \begin{minipage}[t]{0.30\columnwidth}\raggedright
Standard deviation = square root of mean.

Very fast thresholding with table lookup.\strut
\end{minipage} & \begin{minipage}[t]{0.30\columnwidth}\raggedright
Assumes Poisson statistics for the noise.\strut
\end{minipage}\vspace{6pt}\tabularnewline
\begin{minipage}[t]{0.30\columnwidth}\raggedright
Mean + factor * sigma\strut
\end{minipage} & \begin{minipage}[t]{0.30\columnwidth}\raggedright
Accounts for variance differences per pixel.

Handles non-Poisson standard deviations.

Mean/sigma updates track temporal changes\strut
\end{minipage} & \begin{minipage}[t]{0.30\columnwidth}\raggedright
Requires mean \& standard dev. per pixel.

Takes time for mean/sigma filters to spinup.\strut
\end{minipage}\vspace{6pt}\tabularnewline
Maxfilter & Fast thresholding & Only exceedances -- no mean/standard
dev.\vspace{6pt}\tabularnewline
\begin{minipage}[t]{0.30\columnwidth}\raggedright
Maxpixel / Maxframe\strut
\end{minipage} & \begin{minipage}[t]{0.30\columnwidth}\raggedright
Exceedances per frame are natural byproduct

Threshold set by number of frames in a block.

Fast thresholding.\strut
\end{minipage} & \begin{minipage}[t]{0.30\columnwidth}\raggedright
Requires initial pass through a frame block.

Slower if mean/standard dev. estimated.

Mean/sigma must avoid max contamination.

Treats each frame block independently.\strut
\end{minipage}\tabularnewline
\midrule
\multicolumn{3}{c}{\textbf{Meteor Detection -} \emph{Finding propagating line segments}}\tabularnewline
\midrule
\begin{minipage}[t]{0.30\columnwidth}\raggedright
Spatial Change\strut
\end{minipage} & \begin{minipage}[t]{0.30\columnwidth}\raggedright
Match a very simple spatial signature spike.

Useful when no temporal response.\strut
\end{minipage} & \begin{minipage}[t]{0.30\columnwidth}\raggedright
No temporal mitigation of false alarms.\strut
\end{minipage}\vspace{6pt}\tabularnewline
\begin{minipage}[t]{0.30\columnwidth}\raggedright
Clustering \& Tracking\strut
\end{minipage} & \begin{minipage}[t]{0.30\columnwidth}\raggedright
Very fast grouping of exceedances to clusters.

Tracks multiple events at the same time.

Downscaling adds little to runtime costs.

Low latency of detection.\strut
\end{minipage} & \begin{minipage}[t]{0.30\columnwidth}\raggedright
Clustering is more a blob than line detector.

Long streaks need hierarchical down sampling.

De-duplicate multi-level downscaled detects\strut
\end{minipage}\vspace{6pt}\tabularnewline
\begin{minipage}[t]{0.30\columnwidth}\raggedright
Orientation Kernel\strut
\end{minipage} & \begin{minipage}[t]{0.30\columnwidth}\raggedright
Fast convolution of small spatial kernels.

Low latency of detection.\strut
\end{minipage} & \begin{minipage}[t]{0.30\columnwidth}\raggedright
Spatial detections require temporal linkage.

Does not capture cross-track PSF blurring.\strut
\end{minipage}\vspace{6pt}\tabularnewline
\begin{minipage}[t]{0.30\columnwidth}\raggedright
Hough Transform\strut
\end{minipage} & \begin{minipage}[t]{0.30\columnwidth}\raggedright
Detects short and long line segments.

Medium computational load for pixel pairing.

Pixel pairing avoids butterfly self-noise.

Can accumulate Hough space across frames.

Can also use position and orientation (Note 1)

Low latency of detection.\strut
\end{minipage} & \begin{minipage}[t]{0.30\columnwidth}\raggedright
Heavy computation load for classic Hough.

Heavy computation load for phase coded disk.

Heavy computation load for Hueckle xform.\strut
\end{minipage}\vspace{6pt}\tabularnewline
\begin{minipage}[t]{0.30\columnwidth}\raggedright
Matched Filter\strut
\end{minipage} & \begin{minipage}[t]{0.30\columnwidth}\raggedright
Optimal detector in Gaussian noise.

Provides leading edge pick points.

Usable if cued to trim \# of motion hypotheses\strut
\end{minipage} & \begin{minipage}[t]{0.30\columnwidth}\raggedright
Extremely heavy computational load.

Huge latency of detection = non-real-time.

Requires clutter suppression -- whitening.\strut
\end{minipage}\tabularnewline
\bottomrule
\end{longtable}
\end{landscape}

\clearpage
\twocolumn

Note 1 - The Hough space orientation angle can be estimated at an
exceedance pixel by several alternative approaches to the standard
accumulation over all angles Hough transform \citep{Duda1972}. For
example, one can use 1) a phase coded disk applied to a binary
exceedance image \citep{Clode2004}, 2) a Hueckel transform applied to
a gray scale image \citep{Hueckel1973}, 3) pixel pairing of local
neighborhood exceedance pixels (Gural 1999a), or 4) linked exceedances
clustered into line segments with a covariance eigenvalues estimation.

\section{Appendix B: Configuration parameter list}

The following lists the file contents of the configuration parameters used for the
EMCCD detection application.

Read version = 2.11

Sample time/frame in sec = 0.031119

Duty factor = 1.0

Magnitude zero point = 16.72

Interleave type = 0

Backgnd low cutoff (\%) = 5

Backgnd high cutoff (\%) = 50

Number compressed frames M = 64

Max downscaling 1,2,4,8,16 = 2

Primary sigma factor = 0.0

Cell block size (pixels) = 32

Max clusters per frame = 20

Neighbors + self tuplet = 5

Tiny variance = 0.010000

Saturation gray level = 35535.000000

Cluster sigma factor = 1.5

Cluster detection (dB) = -2.0

Tracker FIRM = 3 of 4

Tracker CLOSED = 2 of 3

Tracker TROUBLED = 1

Tracker max tracklets = 200

Tracker max history = 100

MFIL accel model -1,0,+1 = 0

MFIL accel nframes limit1 = 10

MFIL accel nframes limit2 = 15

Detect Vmin (pixels/frame) = 2.000000

Detect Vmax (pixels/frame) = 35.000000

Detect linearity (pixels) = 2.000000

Detect model fit (pixels) = 3.000000

Min \#frames for detect = 3

\#frames to extend track = 8

SNR to cull extended frms = 1.0

PSF size (pixels) = 3

PSF halfwidth (pixels) = 1.130000

Dedupe angle limit (deg) = 10.0

Dedupe rel velocity \% = 40.0

Dedupe max distance (pixels) = 10.0

TRT Threshold = 0.6

MLE Threashold (dB) = 0.0

Reporting option = 28

-\/-\/-\/-\/-\/-\/-\/-\/-\/-\/-\/-\/-\/-\/-\/-\/-\/-\/-\/-\/-\/-\/-\/-\/-\/-\/-\/-\/-\/-\/-\/-\/-\/-\/-\/-\/-\/-\/-\/-\/-\/-\/-\/-\/-\/-\/-\/-\/-\/-\/-\/-\/-\/-\/-\/-\/-\/-\/-\/-\/-\/-\/-\/-\/-\/-\/-\/-\/-\/-\/-\/-

Parameters that can influence sensitivity of detection:
\begin{enumerate}

\item Number of compressed frames (64 = 2.16 sigma; 44 = 2.0 sigma; 32 = 1.87
sigma)

\item Neighbors + self tuplet cluster size (nominally 6)

\item Sigma factor for cluster detection (nominally 2.0)

\item SNR for cluster detection (nominally 0.0)

\item MLE threshold (nominally 0 dB, but 8-10 dB could be used to screen out faint false alarms but also potentially faint meteors)
\end{enumerate}

Reporting option is the sum of the following:

1 = Cluster/tracker detections (no dedupe)

2 = Matched filter detections (no dedupe)

4 = Matched filter detections (de-duplicated)

8 = Matched filter detect BMPs (no dedupe)

16 = ASGARD formatted file of MF detections (deduped)

Max downscaling option to aid blob detection by reducing resolution which includes 1x1, 2x2, 4x4, 8x8, 16x16 up to Max x Max specified

Examples:
\begin{itemize}

\item 1 = 1x1 only (no downscaling single processing path)

\item 2 or 3 = 1x1 and 2x2 downscaling processing paths

\item 4, 5, 6 or 7 = 1x1, 2x2 and 4x4 downscaling processing paths

\item 8 to 15 = 1x1, 2x2, 4x4 and 8x8 downscaling processing paths

\item \textgreater=16 uses 1x1, 2x2, 4x4, 8x8 and 16x16 downscaling paths
\end{itemize}
Match filter acceleration model options for refinement:

-1 = Zero acceleration and zero jerkiness

0 = Constant acceleration and zero jerkiness

1 = Linear acceleration and constant jerkiness

NOTE: This is a default value which gets adjusted down for shorter pick count meteors as follows:

\textless nframes limit1 picks enforces no higher than zero accel

\textless nframes limit2 picks enforces no higher than constant accel

Set limit1 and limit2 to zero to force only one accel model.

\end{appendices}

\bibliography{matched}

\end{document}